\documentclass[prb, floatfix, twocolumn, superscriptaddress]{revtex4-2}
\def\be{\begin{equation}}
\def\ee{\end{equation}}
\def\bea{\begin{eqnarray}}
\def\eea{\end{eqnarray}}
\def\bi{\begin{itemize}}
\def\ei{\end{itemize}}
\usepackage{braket}
\usepackage{xcolor}
\usepackage{graphicx}
\usepackage{amsmath,amstext,amssymb,bm,color,times}
\usepackage[english]{babel}
\usepackage[T1]{fontenc}
\usepackage[utf8]{inputenc}
\usepackage[colorlinks=true,citecolor=blue,linkcolor=magenta]{hyperref}

\begin{document}

%%%%%%%%%%%%%%%%%%%%%%%%%%%%%%%%%%%%%%%%%%%%%%%%%%%%%%%%%%%%%%%%%%%%%%%%%%

\title{
Persistent 
%coherent 
%many-body 
oscillation of a Cooper-pair condensate of topological defects in a nonintegrable quantum Ising chain
}

%%%%%%%%%%%%%%%%%%%%%%%%%%%%%%%%%%%%%%%%%%%%%%%%%%%%%%%%%%%%%%%%%%%%%%%%%%
\author{Francis A. Bayocboc Jr.}
\affiliation{Jagiellonian University, Institute of Theoretical Physics, {\L}ojasiewicza 11, PL-30348 Krak\'ow, Poland}
\author{Jacek Dziarmaga}
\affiliation{Jagiellonian University, Institute of Theoretical Physics, {\L}ojasiewicza 11, PL-30348 Krak\'ow, Poland}
\author{Marek M. Rams}
\affiliation{Jagiellonian University, Institute of Theoretical Physics, {\L}ojasiewicza 11, PL-30348 Krak\'ow, Poland}
\author{Wojciech H. Zurek}
\affiliation{Theory Division, Los Alamos National Laboratory, Los Alamos, New Mexico 87545, USA}

\date{\today}
%%%%%%%%%%%%%%%%%%%%%%%%%%%%%%%%%%%%%%%%%%%%%%%%%%%%%%%%%%%%%%%%%%%%%%%%%%

\begin{abstract}
We identify persistent oscillations in a nonintegrable quantum Ising chain. In the integrable chain with nearest-neighbor interactions, the nature, origin, and decay of post-transition oscillations are tied to the Kibble-Zurek mechanism. Remarkably, when coupling to the next-nearest neighbor is added, the resulting nonintegrable ``zigzag’’ chain (still in the quantum Ising universality class) supports persistent oscillation: Topological defects (kinks) appear as a result of the quantum phase transition. However, in a ``zigzag’’ Ising chain defects can form Cooper pairs. 
The oscillation of the Cooper-pair condensate has a frequency that depends on the binding energy gap between the paired and the unpaired defects, so it can be excited by resonant driving. 
While one might have expected that the integrability-breaking ``zigzag’’ coupling causes relaxation, the oscillations we identify are persistent: Their longevity in our simulations is likely limited only by numerical accuracy. This oscillation of the Cooper-pair condensate of kinks is a manifestation of quantum coherence and should be experimentally accessible. 
\end{abstract}

\maketitle

%%%%%%%%%%%%%%%%%%%%%%%%%%%%%%%%%%%%%%%%%%%%%%%%%%%%%%%%%%%%%%%%%%%%%%%%%%%%%%

%%%%%%%%%%%%%%%%%%%%%%%%%%%%%%%%%%%%%%%%%%%%%%%%%%%%%%%%%%%%%%%%%%%%%%%%%%%%%%

In a recent paper~\cite{transverse_oscillations}, we have identified and characterized oscillations excited by the quench across the critical point of phase transitions in quantum Ising spin systems. These coherent many-body oscillations of the transverse magnetization are caused by quantum superpositions of broken symmetry states left behind in the wake of a nonequilibrium (nonadiabatic) quantum phase transition. 
Superpositions of topological defects---hence, of broken symmetry states---have been considered~\cite{cats_nature_phys} and were recently seen in an experiment~\cite{cats_sci_adv}.
They can be understood as a flagrantly quantum consequence of the Kibble-Zurek mechanism (KZM)~\cite{K-a, *K-b, *K-c,Z-a,*Z-b,*Z-c,*Z-d} which is also responsible for the generation of topological defects in both classical~\cite{KZnum-a,KZnum-b,KZnum-c,KZnum-d,KZnum-e,KZnum-f,KZnum-g,*KZnum-h,*KZnum-i,KZnum-j,KZnum-k,KZnum-l,KZnum-m,that,KZexp-a,KZexp-b,KZexp-c,KZexp-d,KZexp-e,KZexp-f,KZexp-g,QKZexp-a,KZexp-h,KZexp-i,KZexp-j,KZexp-k,KZexp-l,KZexp-m,KZexp-n,KZexp-o,KZexp-p,KZexp-q,lamporesi2013,donadello2016,KZexp-s,KZexp-t,KZexp-u,KZexp-v,KZexp-w,KZexp-x} 
and quantum~\cite{QKZ1,QKZ2,QKZ3,d2005,d2010-a, d2010-b, QKZteor-a,QKZteor-b,QKZteor-c,QKZteor-d,QKZteor-e,QKZteor-f,QKZteor-g,QKZteor-h,QKZteor-i,QKZteor-j,QKZteor-k,QKZteor-l,QKZteor-m,QKZteor-n,QKZteor-o,KZLR1,KZLR2,QKZteor-oo,delcampostatistics,KZLR3,QKZteor-q,QKZteor-r,QKZteor-s,QKZteor-t,sonic,QKZteor-u,QKZteor-v,QKZteor-w,QKZteor-x,roychowdhury2020dynamics,sonic, schmitt2021quantum,RadekNowak,dziarmaga_kinks_2022,transverse_oscillations,QKZexp-a, QKZexp-b, QKZexp-c, QKZexp-d, QKZexp-e, QKZexp-f, QKZexp-g,deMarco2,Lukin18,adolfodwave,2dkzdwave,King_Dwave1d_2022,King_Dwave_glass,king2024computational,miessen2024benchmarking} phase transitions. In particular, frequencies and amplitudes of these postcritical oscillations follow the KZM scalings.

%%%%%%%%%%%%%%%%%%%%%%%%%%%%%%%%%%%%%%%%%%%%%%%%%%%%%%%%%%%%%%%%%%%%%%%%%%%%%%%%%%%
\begin{figure}[t!]
    \includegraphics[width=0.97\columnwidth]{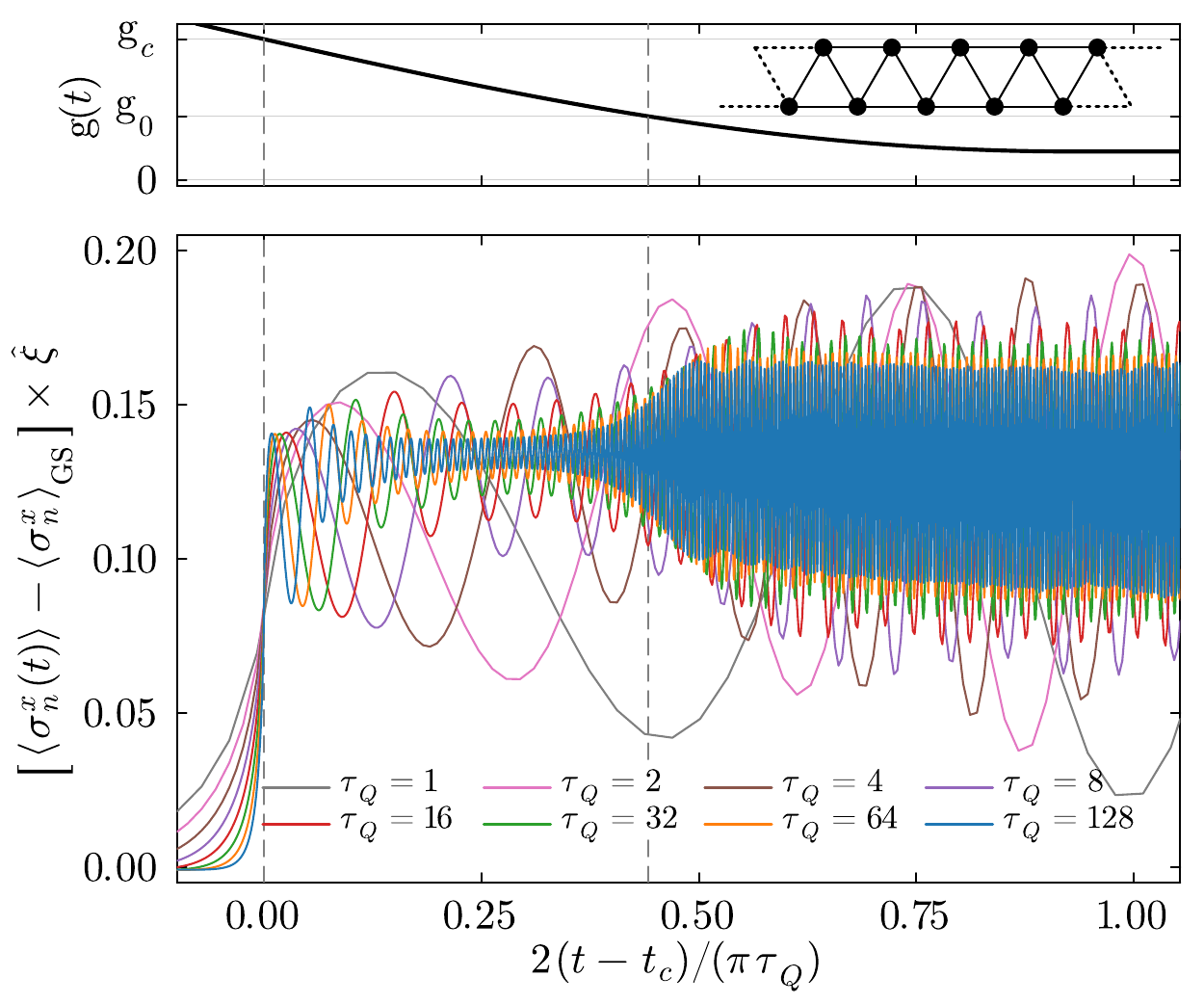}
    \caption{{\bf Persistent coherent oscillation} in the zigzag Ising chain compared and contrasted with the decaying post-KZM oscillations after a quench into the ferromagnetic phase.
    Top:
    The KZ ramp crossing the critical point at $g_c=2.47725$ to end in the ferromagnetic phase at nonzero $g=0.5$.
    The inset shows the zigzag ladder.
    Bottom:
    Rescaled transverse magnetization $\langle\sigma^x_n\rangle - \langle \sigma^x_n \rangle_\mathrm{GS}$ during the ramp as a function of time, where $\langle \sigma^x_n \rangle_\mathrm{GS}$ is its value in the adiabatic ground state and $\hat\xi=\sqrt{\tau_Q}$ is the KZ length. Its oscillation amplitude suddenly increases in the middle of the ferromagnetic phase near $g_0=1.12$. See also the related Fig. S5 in the Supplemental Material \cite{SM}. 
    }
    \label{fig:KZ}
\end{figure}
%%%%%%%%%%%%%%%%%%%%%%%%%%%%%%%%%%%%%%%%%%%%%%%%%%%%%%%%%%%%%%%%%%%%%%%%%%%%%%%%%%%

In Ref.~\onlinecite{transverse_oscillations}, two models were studied. The simplest quantum Ising chain with only nearest-neighbor (NN) couplings is integrable and a paradigmatic example of quantum phase transitions. The Ising chain with both NN and next to NN (NNN) couplings is nonintegrable,
\be
H=-\sum_{n=1}^L \left( g\sigma^x_n + \sigma^z_n\sigma^z_{n+1} + J_{2}\sigma^z_n\sigma^z_{n+2} \right).
\label{eq:Hsigma}
\ee
Here, $\sigma_n^{x,y,z}$ are the Pauli operators on site $n$, $g$ is a transverse field, and $J_{2}$ is the NNN ferromagnetic coupling. It can be implemented with only NN couplings on a zigzag ladder in Fig.~\ref{fig:KZ}.
Both models belong to the same universality class, so their behavior in the immediate vicinity of the critical point is expected to be essentially the same. Indeed, we have demonstrated \cite{transverse_oscillations} that both models exhibit similar oscillatory ``ringdown'' behavior with KZM scalings following the quench.

Dramatic differences in the behavior between the two Ising chains arise away from the transition point. In the integrable quantum Ising chain with only NN couplings, postcritical coherent oscillations gradually decay. By contrast (and to our surprise), in the nonintegrable Ising chain with both NN and NNN couplings, while the post-KZM oscillation still initially decays as in the NN chain, but before disappearing it becomes enhanced, giving rise to oscillation that does not decay on the timescales explored by our numerical simulations. 

These reincarnated oscillations have features that are no longer subject to KZM scalings (see Fig.~\ref{fig:KZ}). In particular, their frequencies are no longer controlled by the size of the Ising chain gap associated with a single quasiparticle (kink).  

The aim of this paper is to investigate properties and to characterize the nature of these oscillations. 
We trace their origin to a bosonic low-energy branch of excitations (kink ``Cooper pairs'') appearing for a weak transverse field. Moreover, we show how they can be efficiently excited without crossing the quantum critical point. 

%%%%%%%%%%%%%%%%%%%%%%%%%%%%%%%%%%%%%%%%%%%%%%%%%%%%%%%%%%%%%%%%%%%%%%%%%%%%%%%%%%%
%\section{Integrable NN chain; ${\bf J_2=0}$}
%\label{sec:NN}
%%%%%%%%%%%%%%%%%%%%%%%%%%%%%%%%%%%%%%%%%%%%%%%%%%%%%%%%%%%%%%%%%%%%%%%%%%%%%%%%%%%
{\bf Integrable NN chain: $\mathbf{J_2=0}$.}
%%%%%%%%%%%%%%%%%%%%%%%%%%%%%%%%%%%%%%%%%%%%%%%%%%%%%%%%%%%%%%%%%%%%%%%%%%%%%%%%%%%
We linearly ramp the transverse field
~\cite{QKZ2,d2005,QKZteor-d,Francuzetal,RadekNowak,dziarmaga_kinks_2022} as
\begin{eqnarray}
    g(t) = g_c \left[ 1 - \epsilon(t) \right] = g_c - g_c (t-t_c) / \tau_Q,
    \label{ramp}
\end{eqnarray}
from a strong field at $t=-\infty$, across the critical point at $g(t_c) = g_c=1$, to $g(t_s)=0$ when the transverse field vanishes and the ramp stops. 
The Jordan-Wigner transformation maps the model to a set of independent Landau-Zener systems that can be solved analytically. In particular, the final density of excited quasiparticles/kinks scales as~\cite{QKZ2,d2005,SM}
\be
\rho = \frac{1}{2\pi\sqrt{2\tau_Q}} \propto \hat\xi^{-1}.
\label{n}
\ee
Here, $\hat\xi$ is the KZ length. The KZM operates within $\pm\hat t$ of $t_c$, where $\hat t\propto\hat\xi^1$.
Prior to the kink count, the final state is a quantum superposition of different numbers~\cite{QKZteor-d,delcampostatistics} and correlated locations of kinks~\cite{roychowdhury2020dynamics, RadekNowak, dziarmaga_kinks_2022} that manifests in coherent oscillations of the transverse magnetization $\sigma^x$~\cite{transverse_oscillations}.

Two new results~\cite{SM} make observation of coherent oscillations more feasible for the present quantum simulation platforms. The first is an analytical formula for transverse magnetization after $t_c+\hat t$,
\be
\delta x = \langle \sigma x \rangle - \langle \sigma x \rangle _ {\rm GS} \approx
2 \rho +
\rho 2 d \frac{57\sqrt{6\pi}}{80} \cos \phi. 
\label{eq:deltax}
\ee
Here, 
$ 
d=(1+f^2)^{-3/4}
$
is a dephasing factor where
\be 
f=\frac{3}{4\pi}\left[ \gamma_E-2\frac{t-t_c}{\tau_Q}+\ln \frac{4(t-t_c)^2}{\tau_Q} \right]
\ee 
with Euler's constant $\gamma_E$, the phase of the oscillations is   
\be 
\phi=\frac{\pi}{4}+2\frac{(t-t_c)^2}{\tau_Q}+\frac32\arctan f,
\ee 
and $\langle \sigma^x \rangle_{\rm GS}$ is the magnetization in the adiabatic ground state. In~\eqref{eq:deltax}, both terms are accurate to their leading order in $\rho$. 
The second result is the NN ferromagnetic correlator,
\bea
\delta^{zz} &=& 
\langle \sigma^z_n\sigma^z_{n+1} \rangle - \langle \sigma^z_n\sigma^z_{n+1} \rangle_{\rm GS} \nonumber\\
&\approx&
-2\rho-g~\rho^2 d \frac{57\sqrt{6\pi}}{80} \cos \phi,
\label{eq:deltazz}
\eea  
which may be more directly accessible in some quantum simulation platforms. 
Together with~\eqref{eq:deltax}, it satisfies
$
-\delta^{zz} - g \delta^x \approx 2(1-g) \rho.
$ 
The excitation energy density on the left-hand side is given by the density of quasiparticles $\rho$ times the (``dressed kink'') quasiparticle gap opening with the distance from the critical point as $2(1-g)$, as predicted by the extended quantum KZM (xQKZM)~\cite{schmitt2021quantum}. 

The oscillations~\eqref{eq:deltax} are damped by dephasing of noninteracting quasiparticles. Breaking the integrability with $J_2=1$, one might expect them to become even more scrambled and attenuated, but, surprisingly, they are reincarnated, as a persistent oscillation with a larger amplitude
%, which we shall now discuss 
(see Fig.~\ref{fig:KZ}). 
%Not only do they become more persistent, but even their amplitude is enhanced.

%%%%%%%%%%%%%%%%%%%%%%%%%%%%%%%%%%%%%%%%%%%%%%%%%%%%%%%%%%%%%%%%%%%%%%%%%%%%%%%%%%%%%%%%%%%%%%%%%
\begin{figure}[t!]
    \includegraphics[width=\columnwidth]{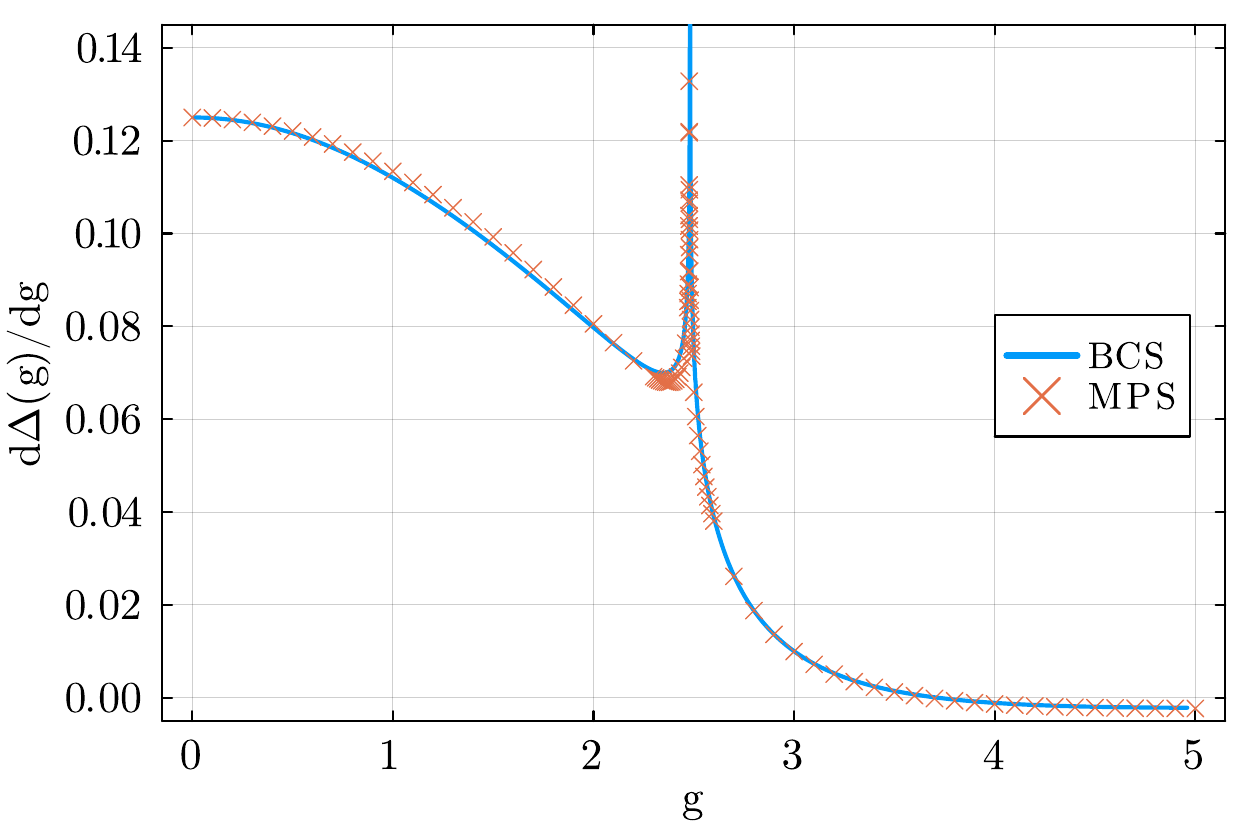}
    \caption{{\bf BCS vs MPS. }
    The first derivative of the gap function $\Delta$ with respect to the field $g$ in the ground state from the BCS theory and matrix product states (MPS). Estimates of $g_{c}$ from BCS and MPS are $g_{c} = 2.48135$ and $g_{c} = 2.47725$, respectively. 
}
    \label{fig:BdG_v_MPS_derivatives}
\end{figure}
%%%%%%%%%%%%%%%%%%%%%%%%%%%%%%%%%%%%%%%%%%%%%%%%%%%%%%%%%%%%%%%%%%%%%%%%%%%%%%%%%%%%%%%%%%%%%%%%%%

%%%%%%%%%%%%%%%%%%%%%%%%%%%%%%%%%%%%%%%%%%%%%%%%%%%%%%%%%%%%%%%%%%%%%%%%%%%%%%%%%%%
%\section{ Kink pair bound states }
%\label{sec:QIM}
%%%%%%%%%%%%%%%%%%%%%%%%%%%%%%%%%%%%%%%%%%%%%%%%%%%%%%%%%%%%%%%%%%%%%%%%%%%%%%%%%%%
{\bf Kink Cooper-pair condensate in a ``zigzag'' Ising chain: $\mathbf{J_2=1}$. }
%%%%%%%%%%%%%%%%%%%%%%%%%%%%%%%%%%%%%%%%%%%%%%%%%%%%%%%%%%%%%%%%%%%%%%%%%%%%%%%%%%%
With a {\it fermionic} operator that annihilates a kink on a bond between sites $n$ and $n+1$~\cite{dziarmaga_kinks_2022},
\be 
f_{n} = \left( \prod_{l\le n} \sigma^x_l \right) \frac{ \sigma^z_{n} - \sigma^z_{n+1}}{2 i} ,
\label{gamma}
\ee
one can rewrite the Hamiltonian~\eqref{eq:Hsigma} with $J_2=1$ as
\bea 
H &=& -2L + 6 \sum_n f_n^\dag f_n -4 \sum_n f_n^\dag f_n f_{n+1}^\dag f_{n+1} \nonumber \\    
& & - g \sum_n \left( f_n^\dag f_{n+1} + f_{n+1} f_n + {\rm h.c.}  \right) .
  \label{eq:Hk}
\eea 
The first term is the energy of the ferromagnetic state, and the second is that of individual kinks in this state. The transverse field in the second line can create/annihilate a pair of kinks on NN bonds by flipping the spin between them. A spin flip can also make a kink hop between NN bonds. The integrability is broken by the quartic attraction between NN kinks.

In the BCS theory, the ground state is assumed to be a fermionic Gaussian state $\ket{0}$. After a Fourier transform
\be 
f_n = e^{i\pi/4} \sum_k \frac{e^{ik(n+1/2)}}{\sqrt{L}} f_k, 
\label{eq:Fourier}
\ee
and a Bogoliubov transformation,
\be 
f_k = u_k \gamma_k + v_{-k}^* \gamma_{-k}^\dag,
\label{eq:kBog}
\ee 
that introduces fermionic quasiparticles annihilated (created) by $\gamma^{(\dag)}$, the energy is minimized by a vacuum,
$
\gamma_k \ket{0} = 0 ,
$
where $(u_k,v_k)$ solve the Bogoliubov-de Gennes equations:
\be 
\omega_{k}
\left(
\begin{array}{c}
u_k \\
v_k
\end{array}
\right) =
\left(
\begin{array}{cc}
H_k   & D_k \\
D^*_k & -H_k
\end{array}
\right)
\left(
\begin{array}{c}
u_k \\
v_k
\end{array}
\right).
\ee 
Here, $H_k=2(3-4\rho)-2(g-4t_f)\cos k$, $D_k=-2(g+4\Delta)\sin k$, and the quasiparticle dispersion $\omega_k=\sqrt{H_k^2+D_k^2}$. Here, $\rho=\bra{0} f_n^\dag f_n \ket{0}$ is density of kinks, $\Delta=\bra{0} f_{n+1} f_n \ket{0}$ is their anomalous density, and $t_f=\bra{0} f_{n+1}^\dag f_n \ket{0}$ is a correction to the kink hopping rate. 
The BCS solution compares well with the matrix product states (MPS)~\cite{Vanderstraeten2019} (see Fig.~\ref{fig:BdG_v_MPS_derivatives} and Figs. S1 and S2 in the Supplemental Material \cite{SM}). The critical $g_c=2.48135$ by BCS lies within $0.2\%$ of $g_c=2.47725$ obtained by MPS algorithm. 
The quasiparticle (dressed kink) dispersion is
\bea 
\omega_k & = & 
\omega_\gamma - 2 t_\gamma \cos k - 2 t'_\gamma \cos 2k + {\cal O}(g^3),
\label{eq:omegap}
\eea 
where $\omega_\gamma=6+g^2/8$, $t_\gamma=g$, and $t'_\gamma=3g^2/16$ (see Supplemental Material \cite{SM}).

%%%%%%%%%%%%%%%%%%%%%%%%%%%%%%%%%%%%%%%%%%%%%%%%%%%%%%%%%%%%%%%%%%%%%%%%%%%%%%%%%%%%%%%%%%%
\begin{figure}[t!]
    \includegraphics[width=\columnwidth]{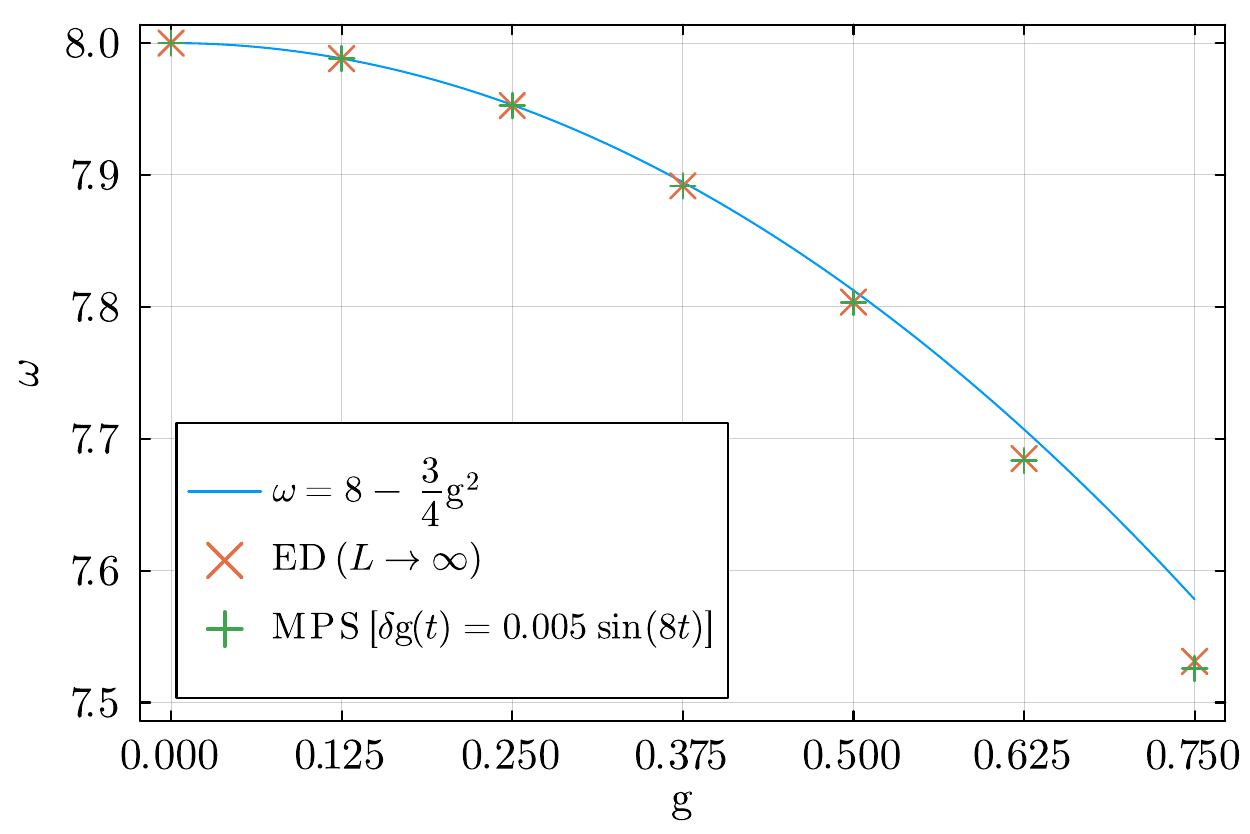}
    \caption{{\bf Oscillation frequency. }    
    The plot compares~\eqref{eq:omega} with the gap between the ground and first excited state obtained with exact diagonalization (ED) extrapolated to infinite periodic chain size ($L\to\infty$), and transverse oscillation frequency 
    of the Cooper-pair condensate of topological defects
    after the periodic drive~\eqref{eq:delta_H} with a small amplitude obtained by a fit to MPS simulations. The ED and MPS error bars are on the order of the marker stroke width.
    }
    \label{fig:dispersion}
\end{figure}
%%%%%%%%%%%%%%%%%%%%%%%%%%%%%%%%%%%%%%%%%%%%%%%%%%%%%%%%%%%%%%%%%%%%%%%%%%%%%%%%%%%%%%%%%%%

The quartic attraction in \eqref{eq:Hk} motivates the introduction of NN pair annihilation operators 
\be
b_n=\gamma_{n+1}\gamma_{n}.
\label{eq:b}
\ee
They are approximately bosonic at a low pair density where an effective Hamiltonian can be obtained as~\cite{SM} 
\bea 
H_b &=&
\omega_b\sum_n b_n^\dag b_n
- t_b \sum_n\left(b^\dag_{n}b_{n+1}+{\rm h.c.}\right) \nonumber\\
& &
- t'_b    \sum_n\left(b^\dag_{n}b_{n+2}+{\rm h.c.}\right) + {\cal O}\left( g^3 \right).
\label{eq:Heffp}
\eea
Here, $\omega_b=8-g^2/4$, $t_b=g^2/8$, and $t'_b=g^2/8$. A pair has minimal energy for zero pair quasimomentum,
\bea 
\omega = \omega_b - 2t_b - 2t'_b = 8-\frac34 g^2,
\label{eq:omega}
\eea 
which agrees with the gap from exact diagonalization in Fig.~\ref{fig:dispersion}.
The Hamiltonian \eqref{eq:Heffp} becomes accurate when the pairs are the lowest branch of excitations, i.e., the pair gap in~\eqref{eq:omega} is less than twice the minimal quasiparticle ``dressed kink'' gap in~\eqref{eq:omegap}, $\omega<2\omega_0$, below $g_0\approx1.12$. The terms neglected in \eqref{eq:Heffp}, that are breaking kink pairs or changing the number of fermions, are suppressed by the pair gap and the gap between the pair branch and the fermionic quasiparticles---dressed kinks. 

Within a dynamic BCS theory, the KZ ramp results in a fermionic Gaussian state where pairs of $\gamma$ quasiparticles are not bound but free to move along the chain~\cite{dziarmaga_kinks_2022}. Bound pairs would not affect the long-range order, while the actual state has finite ferromagnetic domains of typical size $\hat\xi$. 
The free quasiparticles occupy quasimomenta up to $\propto\hat\xi^{-1}$. Their nontrivial dispersion results in different quasimomenta, contributing a continuum of different frequencies to the oscillations. Inevitably, this forces them to go out of phase (dephase), thus suppressing their amplitude.

\begin{figure}[t!]
    \includegraphics[width=\columnwidth]{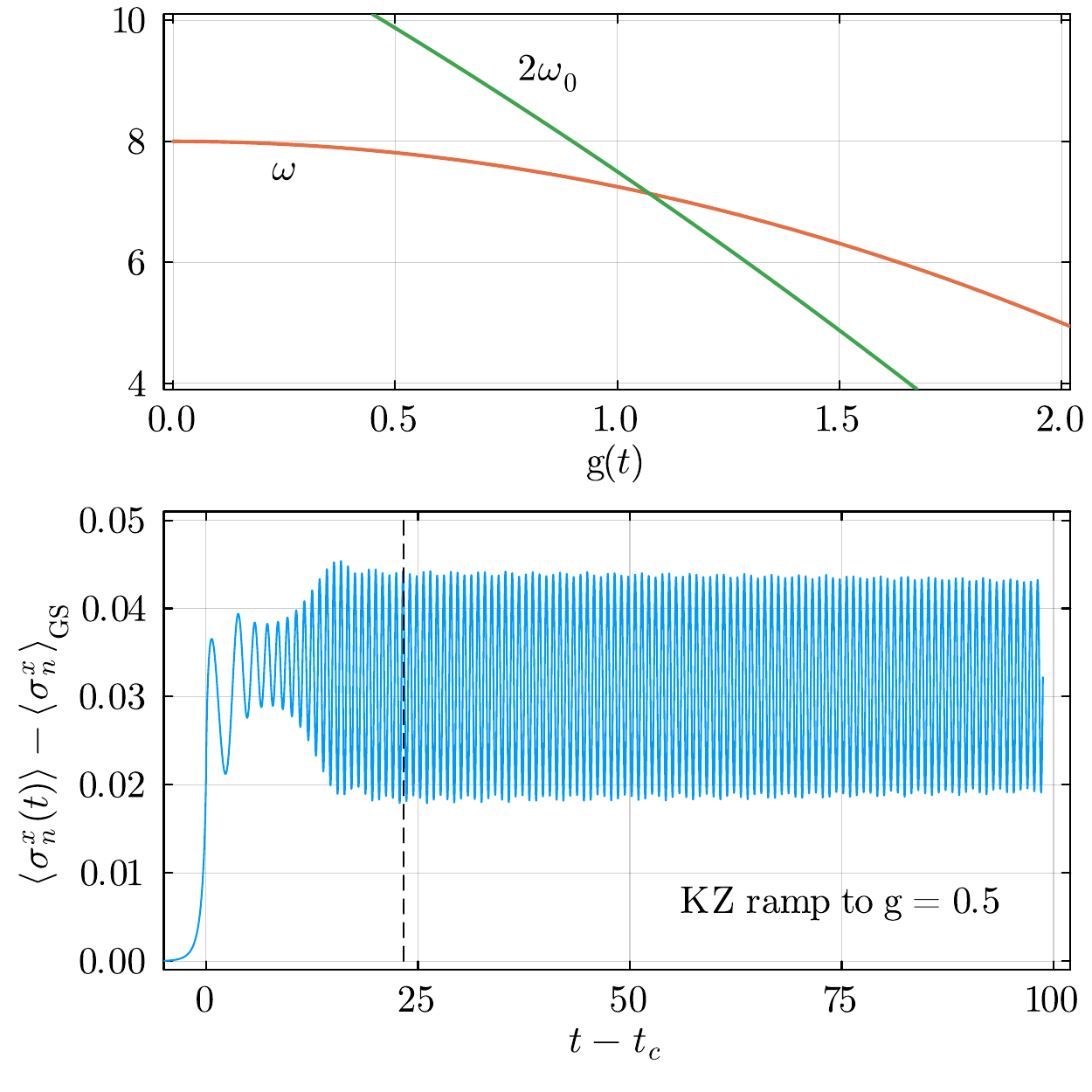}
    \caption{{\bf Oscillations of the Cooper-pair condensate of topological defects after KZ ramp. }
    Top: 
    The energy gap for creating a pair of fermionic quasiparticles, $2\omega_0$, vs that for their bound pair, $\omega$. They crossover near $g_0=1.12$. 
    Bottom:
    Oscillations after a smooth KZ ramp to $g=0.5$ with $\tau_Q=16$, using a smooth ramp $g(t) = g_{c}\{2 - [1-0.5/(2g_{c})][1+\sin(t/\tau_{Q})]\}, t\in[-\tau_{Q}\times\pi/2,\tau_{Q}\times\pi/2]$. The dashed line indicates the end of the ramp, followed by a free evolution at $g=0.5$. 
    A quality factor Q (a decay time divided by a period of oscillation) during the free evolution increases from $Q{\approx}4012$ for $\tau_Q=16$ to $Q\approx 8000$ for $\tau_Q=32$. The estimates of Q here and below may well be significantly affected by numerical imperfections.
    }
    \label{fig:oscillations}
\end{figure}

However, as $g$ is slowly ramped further down below $g_0$, the individual quasiparticles become unstable towards the lower energy branch of pair excitations (see Fig.~\ref{fig:oscillations}).
The pair dispersion is weaker than the dispersion of quasiparticles [$\propto g^2$ in~\eqref{eq:Heffp} versus $\propto g$ in~\eqref{eq:omegap}, respectively] but not weak enough to explain the absence of dephasing that we observe for times much longer than $\propto g^{-2}$. 
The pair dispersion is made irrelevant by a Bose-enhanced transfer of quasiparticles into pairs with zero quasimomentum. Unlike fermions, the bosonic kink pairs can occupy the same quasimomentum state where they all have the same oscillation frequency
of the Cooper-pair condensate of topological defects. 
The appearance of pairs manifests in Fig.~\ref{fig:KZ} by the amplification of oscillations near $g_0$, as can be explained with a simple example: For $\ket{\psi}=\alpha \ket{0} + \beta b_{n-1}^\dag \ket{0} e^{-i\omega t}$ we obtain
\bea 
&& 
\bra{\psi} \sigma^x_{n} \ket{\psi}\approx \nonumber\\
&&
\bra{\psi}  
\gamma_{n-1}^\dag \gamma_{n} + \gamma_{n}^\dag \gamma_{n-1} 
- \gamma_{n-1} \gamma_{n} - \gamma_{n}^\dag \gamma_{n-1}^\dag 
\ket{\psi} = \nonumber\\
&&
\alpha^*\beta e^{-i\omega t} + \alpha\beta^* e^{+i\omega t}.
\label{eq:example}
\eea 
As the pair is a reversed $n$th spin (dressed in quantum fluctuations), it manifests in oscillations of the spin-reversing $\sigma^x_n$. 

%%%%%%%%%%%%%%%%%%%%%%%%%%%%%%%%%%%%%%%%%%%%%%%%%%%%%%%%%%%%%%%%%%%%%%%%%%%%%%%%%%%%%%%%%%%
%%%%%%%%%%%%%%%%%%%%%%%%%%%%%%%%%%%%%%%%%%%%%%%%%%%%%%%%%%%%%%%%%%%%%%%%%%%%%%%%%%%%%%%%%%%

%%%%%%%%%%%%%%%%%%%%%%%%%%%%%%%%%%%%%%%%%%%%%%%%%%%%%%%%%%%%%%%%%%%%%%%%%%%%%%%%%%%%%%%%%%%
\begin{figure}[t!]
    \includegraphics[width=\columnwidth]{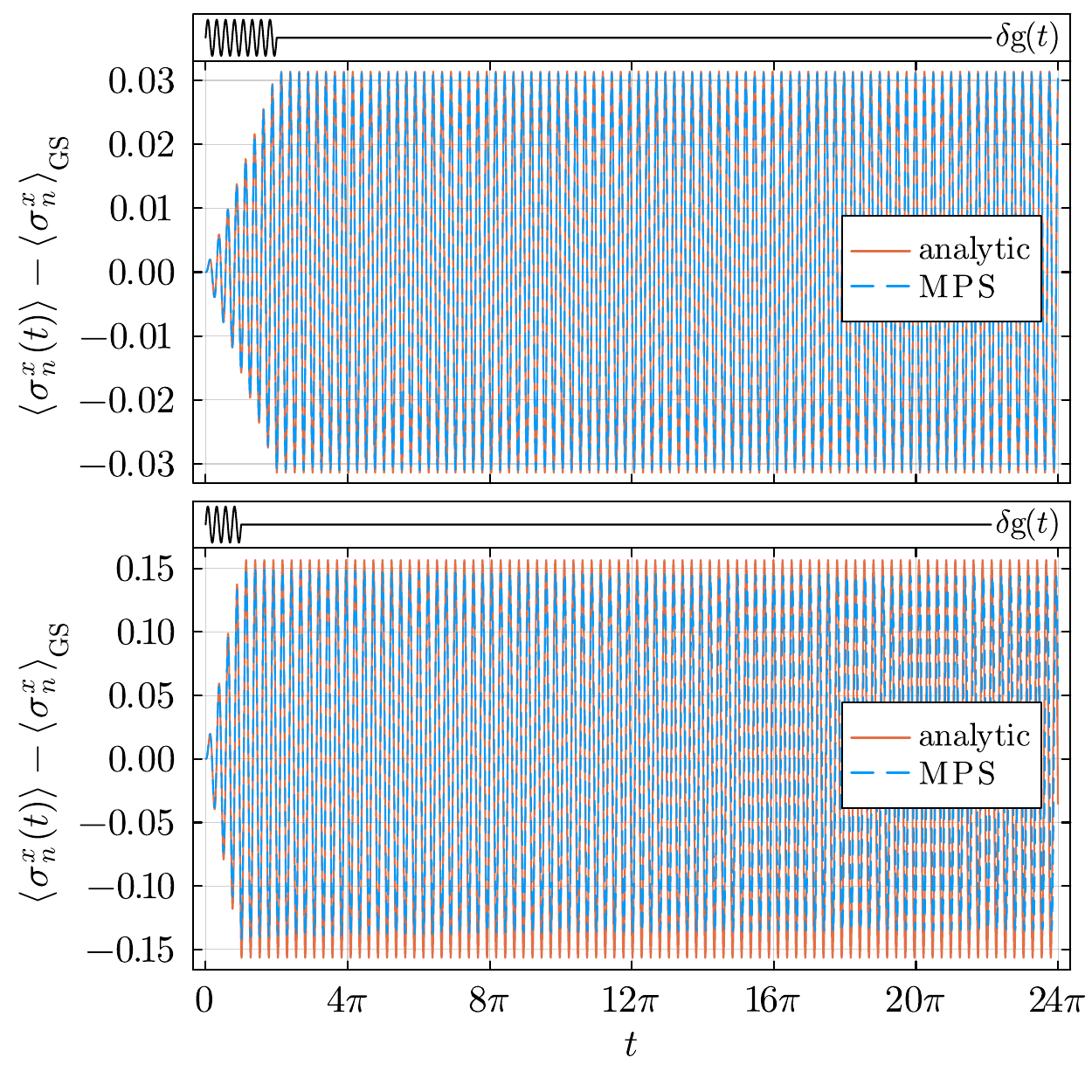}
    \caption{{\bf Oscillations of the Cooper-pair condensate of topological defects after periodic driving. }
    Top: 
    Coherent transverse field oscillations at $g=0.25$ after driving $\delta g(t)=0.005\sin(8t)$ close to resonance for time $2\pi$. The plot compares~\eqref{eq:mx_time} with MPS simulations. The maximum percentage difference between~\eqref{eq:mx_time} and MPS simulations is less than $0.8\%$.
    Bottom: 
    The same at $g=0.5$ after driving $\delta g(t)=0.05\sin(8t)$ for time $\pi$.
    The Q factors of the top and bottom plots are $Q\approx 461504$ and $Q\approx 8307$, respectively.
    }
    \label{fig:periodic}
\end{figure}
%%%%%%%%%%%%%%%%%%%%%%%%%%%%%%%%%%%%%%%%%%%%%%%%%%%%%%%%%%%%%%%%%%%%%%%%%%%%%%%%%%%%%%%%%%%

%%%%%%%%%%%%%%%%%%%%%%%%%%%%%%%%%%%%%%%%%%%%%%%%%%%%%%%%%%%%%%%%%%%%%%%%%%%%%%%%%%%%%%%%%%%
% \section{Periodic driving}
% \label{sec:periodic}
%%%%%%%%%%%%%%%%%%%%%%%%%%%%%%%%%%%%%%%%%%%%%%%%%%%%%%%%%%%%%%%%%%%%%%%%%%%%%%%%%%%%%%%%%%%
{\bf Periodic drive. }
%%%%%%%%%%%%%%%%%%%%%%%%%%%%%%%%%%%%%%%%%%%%%%%%%%%%%%%%%%%%%%%%%%%%%%%%%%%%%%%%%%%%%%%%%%%
The KZ ramp is not the most efficient way to excite coherent oscillations of the Cooper-pair condensate of topological defects. Starting from a ground state at $g$, the pair condensate can also be readily excited by a time-dependent perturbation to the Hamiltonian
\be 
\delta H = \delta g(t) \sum_n \sigma^x_n  .
\label{eq:delta_H}
\ee
Assuming a small density of pairs, where their hard-core nature can be ignored, and there is a gap between paired and unpaired quasiparticles ($g\ll g_0$), we can approximate the transverse magnetization as
\be 
\sum_n \left( \sigma^x_n - \langle \sigma^x_n \rangle_{GS} \right) \approx
\sum_n \left( b_n + b_n^\dag \right) = 
\sqrt{L} \left(\tilde b_{0}+\tilde b_{0}^\dag\right),
\label{eq:X_trunc}
\ee
where $\tilde b_k=e^{i\pi/4}b_k$, and the pair Hamiltonian~\eqref{eq:Heffp} becomes
\bea 
H'_b & = &
\sum_k \omega_{b,k} \tilde b_k^\dag \tilde b_k + 
\delta H.
\eea
Here, $\omega_{b,k}=\omega_b - 2t_b\cos k - 2t'_b\cos 2k$. 
Since only $k=0$ is driven by $\delta H$ in~\eqref{eq:X_trunc}, it reduces to a driven oscillator,
\bea 
H'_{\rm b,0} &=&
\omega 
\tilde b_0^\dag \tilde b_0 +
\delta g(t) \sqrt{L} \left(\tilde b_0+\tilde b_0^\dag\right).
\eea
The driving generates a coherent state of $\tilde b_0$ bosons. Starting from the ground state, the transverse magnetization evolves as
\be 
\left( \frac{d^2}{dt^2} + \omega^2 \right) 
\left(
\langle \sigma^x_n \rangle - \langle \sigma^x_n \rangle_{GS}
\right) =
-2\omega~ \delta g(t).
\label{eq:mx_time}
\ee 
The pairs manifest themselves also in the correlators,
\bea 
&& 
- \sum_n \langle \sigma^z_n \sigma^z_{n+1} \rangle \approx 
- \sum_n \langle \sigma^z_n \sigma^z_{n+2} \rangle \approx \nonumber\\
&&
~~~~~~~~~~
-L + 4 \sum_n \langle b_n^\dag b_n \rangle +
\frac{g}{2} \sqrt{L} \left\langle \tilde b_0 + \tilde b_0^\dag \right\rangle . 
\label{eq:NN_NNN_osc}
\eea 
A weak driving for a finite time results in persistent oscillations with a well-defined frequency. Figure~\ref{fig:periodic} shows their examples for $g=0.25,0.5$. Figure~\ref{fig:dispersion} demonstrates the frequency to be consistent with the gap estimate from exact diagonalization and the perturbative formula~\eqref{eq:omega}. 

%%%%%%%%%%%%%%%%%%%%%%%%%%%%%%%%%%%%%%%%%%%%%%%%%%%%%%%%%%%%%%%%%%%%%%%%%%%%%%%%%%%%%%%%%%%
% \section{Conclusion}
% \label{sec:conclusion}
%%%%%%%%%%%%%%%%%%%%%%%%%%%%%%%%%%%%%%%%%%%%%%%%%%%%%%%%%%%%%%%%%%%%%%%%%%%%%%%%%%%%%%%%%%%
{\bf Conclusion. }
%%%%%%%%%%%%%%%%%%%%%%%%%%%%%%%%%%%%%%%%%%%%%%%%%%%%%%%%%%%%%%%%%%%%%%%%%%%%%%%%%%%%%%%%%%%
In the ``zigzag’’ quantum Ising chain, individual topological defects (dressed kinks---fermionic Bogoliubov quasiparticles that appear below the critical transverse magnetic field) give rise to tightly bound Cooper pairs as the lowest branch of excitations.

The oscillation of the condensate of Cooper pairs of kinks can be excited either by the KZ ramp into the ferromagnetic broken symmetry state or by periodic driving. Despite the nonintegrability of the ``zigzag'' Ising chain, the oscillation of the Cooper-pair condensate of topological defects is persistent---once excited, either by the quench or by resonant driving, it does not appear to decay. 

There are other phenomena (such as quantum many-body scars~\cite{qmb_scars_rydberg,qmb_scars_eth,qmb_scars_rydberg51,scars_rydberg} or quantum time crystals~\cite{Wilczek_TimeCrystals_2012,Sacha_review}) where oscillation is seen or expected. 
 What is unique in the ``zigzag''
Ising chain is that the oscillation is both a manifestation of Cooper pairing of topological defects and of quantum coherence of the resulting BCS-like condensate.

\vspace{9pt}
The data used for the figures in this article are openly available from the RODBUK repository at Ref.~\onlinecite{UJ/4NB7IE_2025}.

%%%%%%%%%%%%%%%%%%%%%%%%%%%%%%%%%%%%%%%%%%%%%%%%%%%%%%%%%%%%%%%%%%%%%%%%%%%%% 
\acknowledgements
%%%%%%%%%%%%%%%%%%%%%%%%%%%%%%%%%%%%%%%%%%%%%%%%%%%%%%%%%%%%%%%%%%%%%%%%%%%%%
%{\bf Acknowledgements.}
%
This research was 
funded by the National Science Centre (NCN), Poland, under project 2021/03/Y/ST2/00184 within the QuantERA II Programme that has received funding from the European Union’s Horizon 2020 research and innovation programme under Grant Agreement No 101017733 (FB),
NCN under projects 2019/35/B/ST3/01028 (JD) and 2020/38/E/ST3/00150 (MMR) 
, and
Department of Energy under the Los Alamos National Laboratory LDRD Program (WHZ).
The research was also supported by a grant from the Priority Research Area DigiWorld under the Strategic Programme Excellence Initiative at Jagiellonian University (JD, MMR).

%%%%%%%%%%%%%%%%%%%%%%%%%%%%%%%%%%%%%%%%%%%%%%%%%%%%%%%%%%%%%%%%%%%%%%%%%%%%

%%%%%%%%%%%%%%%%%%%%%%%%%%%%%%%%%%%%%%%%%%%%%%%%%%%%%%%%%%%%%%%%%%%%%%%%%%%%
\bibliographystyle{apsrev4-2}
\bibliography{KZref.bib}

%%%%%%%%%%%%%%%%%%%%%%%%%%%%%%%%%%%%%%%%%%%%%%%%%%%%%%%%%%%%%%%%%%%%%%%%%%%%%%%%%%%%%%%%%%%%%%%%%%%%%%%%%%%%%%
\end{document}

%%%%%%%%%%%%%%%%%%%%%%%%%%%%%%%%%%%%%%%%%%%%%%%%%%%%%%%%%%%%%%%%%%%%%%%%%%

\title{
Supplemental Material: Persistent oscillation of Cooper pair condensate of topological defects in a nonintegrable quantum Ising chain
}

%%%%%%%%%%%%%%%%%%%%%%%%%%%%%%%%%%%%%%%%%%%%%%%%%%%%%%%%%%%%%%%%%%%%%%%%%%
\author{Francis A. Bayocboc Jr.}
\affiliation{Jagiellonian University, Institute of Theoretical Physics, {\L}ojasiewicza 11, PL-30348 Krak\'ow, Poland}
\author{Jacek Dziarmaga}
\affiliation{Jagiellonian University, Institute of Theoretical Physics, {\L}ojasiewicza 11, PL-30348 Krak\'ow, Poland}
\author{Marek M. Rams}
\affiliation{Jagiellonian University, Institute of Theoretical Physics, {\L}ojasiewicza 11, PL-30348 Krak\'ow, Poland}
\author{Wojciech H. Zurek}
\affiliation{Theory Division, Los Alamos National Laboratory, Los Alamos, New Mexico 87545, USA}

\date{\today}
%%%%%%%%%%%%%%%%%%%%%%%%%%%%%%%%%%%%%%%%%%%%%%%%%%%%%%%%%%%%%%%%%%%%%%%%%%

\maketitle

%%%%%%%%%%%%%%%%%%%%%%%%%%%%%%%%%%%%%%%%%%%%%%%%%%%%%%%%%%%%%%%%%%%%%%%%%%%%%%

%%%%%%%%%%%%%%%%%%%%%%%%%%%%%%%%%%%%%%%%%%%%%%%%%%%%%%%%%%%%%%%%%%%%%%%%%%%%%%

The supplementary material can be divided into two parts. The first one, from Sec.~\ref{app:QIM} to Sec.~\ref{app:transoscduring}, outlines the complete solution of the integrable NN quantum Ising chain that naturally leads to the new results presented in the main text. The standard parts of the solution are based on four papers: two older items~\cite{d2005,QKZteor-e} and three more recent ones~\cite{RadekNowak,dziarmaga_kinks_2022,transverse_oscillations}. They are included here to make the supplementary material self-contained. Sec.~\ref{app:linearquench} derives oscillations between $t_c+\hat t$ and the end of the ramp, and Sec.~\ref{app:transoscduring} estimates their dephasing rate and the quality factor. The second part, from~\ref{app:BCS} to~\ref{app:amplitude}, deals with the nonintegrable chain. 

Both the first part and the BCS theory in Sec.~\ref{app:BCS} use a fermionic Bogoliubov theory. The two theories are not the same and should not be confused even though we use the same symbols like $u_k,v_k$. The former is in the representation of Jordan-Wigner fermions, while the latter is in that of kinks. 

%%%%%%%%%%%%%%%%%%%%%%%%%%%%%%%%%%%%%%%%%%%%%%%%%%%%%%%%%%%%%%%%%%%%%%%%%%%%%%%%%%%
\section{ Quantum Ising chain }
\label{app:QIM}
%%%%%%%%%%%%%%%%%%%%%%%%%%%%%%%%%%%%%%%%%%%%%%%%%%%%%%%%%%%%%%%%%%%%%%%%%%%%%%%%%%%

The Hamiltonian for transverse field quantum Ising chain reads
\be
H~=~-\sum_{n=1}^L \left(  \sigma^z_n\sigma^z_{n+1}  + g\sigma^x_n\right)~,
\label{Hsigma}
\ee
where we consider a system of $L$ spins one-half with periodic boundary conditions, $ \vec\sigma_{L+1}~=~\vec\sigma_1$.  In the limit of $L\to\infty$, there are quantum critical points at $g_c=\pm1$, respectively, that separate the paramagnetic phase for $|g|>1$ from the ferromagnetic phase for $|g|<1$. For simplicity of presentation, we additionally assume that $L$ is even. The Jordan-Wigner transformation,
\bea
&&
\sigma^x_n~=~1-2 c^\dagger_n  c_n~, \\
&&
\sigma^z_n~=~
-\left( c_n + c_n^\dagger\right)
 \prod_{m<n}(1-2 c^\dagger_m c_m)~,
 \label{JW}
\eea
introduces fermionic annihilation ($c_n$) and creation ($c^\dagger_n$) operators. It maps the Hamiltonian in Eq.~\eqref{Hsigma} to
\be
 H~=~P^+~H^+~P^+~+~P^-~H^-~P^-~.
\label{Hc}
\ee
The projectors on subspaces with even ($+$) and odd ($-$) numbers of $c$-quasiparticles read
\be
P^{\pm}=
\frac12\left[1\pm\prod_{n=1}^L\sigma^x_n\right]=
\frac12\left[1~\pm~\prod_{n=1}^L\left(1-2c_n^\dagger c_n\right)\right].
\label{Ppm}
\ee
The reduced Hamiltonians in each parity subspace,
\be
H^{\pm}~=~
\sum_{n=1}^L
\left[
g \left(c_n^\dagger  c_n -\frac12\right) - c_n^\dagger  c_{n+1} + c_n  c_{n+1} \right]  +{\rm h.c.},
\label{Hpm}
\ee
differ in boundary conditions. Namely, in $H^-$ we assume periodic boundary conditions, $c_{L+1}=c_1$, and in $H^+$ we have antiperiodic boundary conditions, $c_{L+1}=-c_1$. 

The parity of the number of $c$-quasiparticles commutes with the Hamiltonian. As the ground state for  $g\gg1$ has even parity, we limit ourselves to that relevant subspace. The next step in the diagonalization of $H^+$ is a Fourier transform, 
\be
c_n~=~ 
\frac{e^{-i\pi/4}}{\sqrt{L}}
\sum_k c_k e^{ikn}~,
\label{Fourier}
\ee
with half-integer pseudo-momenta consistent with the antiperiodic boundary conditions,
\be
k~=~
\pm \frac12 \frac{2\pi}{L},
\pm \frac32 \frac{2\pi}{L},
\dots,
\pm \frac{L-1}{2} \frac{2\pi}{L}~.
\label{k}
\ee
After this transformation, the Hamiltonian takes the form
\bea
H^+~=
&\sum_k&
\left[
(g-\cos k) \left(c_k^\dagger c_k - c_{-k} c_{-k}^\dagger \right)+ \right. \nonumber\\
&& 
\left.~~
\sin k
 \left(
 c^\dagger_k c^\dagger_{-k}+
 c_{-k} c_k
\right)
\right]~.
\label{Hck}
\eea
Its diagonalization is completed by a Bogoliubov transformation,
\be
c_k~=~
u_k  \gamma_k + v_{-k}^*  \gamma^\dagger_{-k}~,
\label{Bog}
\ee
where the Bogoliubov modes $(u_k,v_k)$ follow as eigenstates of the Bogoliubov-de Gennes equations,
\bea
\epsilon~ u_k &=& +2(g-\cos k) u_k+2\sin k~ v_k,\nonumber\\
\epsilon~ v_k &=& -2(g-\cos k) v_k+2\sin k~ u_k. \label{stBdG}
\eea
There are two eigenstates for each value of $k$, with eigenfrequencies $\epsilon=\pm\epsilon_k$, 
\be
\epsilon_k~=~2\sqrt{(g-\cos k)^2+\sin^2 k}~.
\label{epsilonk}
\ee
The eigenstates with positive frequency, $(u^+_k,v^+_k) = (\cos(\theta_k/2), \sin(\theta_k/2))$,
define a fermionic quasiparticle operator $\gamma_k~=~u_k^{+*} c_k + v^+_{-k} c_{-k}^\dagger$, where angles $\theta_k$ satisfy
$(\cos \theta_k, \sin \theta_k) = \frac{2}{\epsilon_k} (g-\cos k, \sin k)$. The negative frequency ones, with 
$(u^-_k,v^-_k)=(v^+_k,-u^+_k)$, formally define $\gamma_k^-=u_k^{-*}c_k+v_{-k}^-c_{-k}^\dagger=-\gamma_{-k}^\dagger$. 
After the Bogoliubov transformation, the Hamiltonian reads
\be
H^+~=~
\sum_k \epsilon_k~
\left( \gamma_k^\dagger \gamma_k~ - \frac12 \right).
\label{Hgamma}
\ee
Note that, due to the projector $P^+$ in Eq.~\eqref{Hc}, only states with even numbers of $c$-quasiparticles belong to the spectrum of $H$. 

The quasiparticle dispersion in Eq.~\eqref{epsilonk} implies a linear dispersion for small $k$ at the critical $g=1$, $\epsilon_k\approx 2|k|$, and the dynamical exponent $z$ is equal to $1$. Moreover, for $k=0$, we have $\epsilon_0\propto |g-1|^1$ and
$z\nu=1$. Finally, the correlation length exponent $\nu$ is equal to $1$. 

%%%%%%%%%%%%%%%%%%%%%%%%%%%%%%%%%%%%%%%%%%%%%%%%%%%%%%%%%%%%%%%%%%%%%
\section{ Linear quench and the Landau-Zener problem}
\label{app:linearquench}
%%%%%%%%%%%%%%%%%%%%%%%%%%%%%%%%%%%%%%%%%%%%%%%%%%%%%%%%%%%%%%%%%%%%%

The Hamiltonian follows a linear ramp in the transverse field,
\be
g(t\leq0)~=~-\frac{t}{\tau_Q},
\label{linear}
\ee
with the quench rate $\tau_Q$. For convenience, here we fix the time when the ramp reaches $g(t_s)=0$ at $t_s=0$ (we use $t_s$ in the main text for clarity). As such, time $t$ runs from $-\infty$ to $0$ when the ramp stops at transverse field $g=0$, crossing the critical point at $g_c=1$ when $t_c=-\tau_Q$. The system starts in the ground state at $g\to\infty$, where $(u_k, v_k) = (1, 0)$, and is the vacuum state annihilated by all corresponding Bogoliubov operators, $\gamma_k |0\rangle = 0$. 

In addressing the dynamical problem, it is convenient to employ the Heisenberg picture. The state of the system stays as the vacuum of Bogoliubov operators $\gamma_k$, while the Bogoliubov modes evolve according to the Heisenberg equation of motion $i\frac{d}{dt}c_k=[c_k,H^+]$. Following the time-dependent Bogoliubov transformation,
\be
c_k = u_k(t)  \gamma_k + v_{-k}^*(t)  \gamma^\dagger_{-k},
\label{tildeBog}
\ee
this gives time-dependent Bogoliubov-de Gennes equations,
\bea
i\frac{d}{dt} u_k &=&
+2\left(g(t)-\cos k\right) u_k +
 2 \sin k~ v_k~,\nonumber\\
i\frac{d}{dt} v_k &=&
-2\left(g(t)-\cos k\right) v_k +
 2 \sin k~ u_k~,
\label{dynBdG} 
\eea
and the initial condition is $(u_k(-\infty),v_k(-\infty))=(1,0)$. 

Introducing a new time variable,
\be
\tau~=~4\tau_Q\sin k\left(\frac{t}{\tau_Q}+\cos k\right),
\label{tau}
\ee
that runs from $-\infty$ to $\tau^{\rm final}_k=2\tau_Q\sin(2k)$ for $t=0$,
allows one to map Eq.~\eqref{dynBdG} to the Landau-Zener (LZ) problem~\cite{d2005,QKZteor-e},
\bea
i\frac{d}{d\tau} u_k &=&
-\frac12 \tau\Delta_k~ u_k + \frac12 v_k,\nonumber\\
i\frac{d}{d\tau} v_k &=&
+\frac12 \tau\Delta_k~ v_k + \frac12 u_k.\label{LZ}
\label{BdGLZ}
\eea
Here, $\Delta_k=(4\tau_Q\sin^2 k)^{-1}$ sets an efficient rate of the transition for given $k$. 

Only modes with small $k$ that have small energy gaps at their anti-crossing point can get excited when the ramp is slow.
For such modes, $\tau^{\rm final}_k$ is much longer than the time when the anti-crossing is completed, and we are allowed to use the LZ formula,
\be
p_k~\approx~
e^{-\frac{\pi}{2\Delta_k}}~\approx~
e^{-2\pi\tau_Qk^2}~,
\label{LZpk}
\ee 
where approximations become accurate for $\tau_Q\gg1$. 
Eq.~\eqref{LZpk} gives the probability that a pair of quasiparticles with quasimomenta $+k$ and $-k$ got excited.
The mean density of kinks at $g=0$ is simply given by $\rho = \sum_k p_k / L$~\cite{d2005}. Taking the limit $L\to\infty$,
\be
\rho=\lim_{L\to\infty}\frac{1}{L}\sum_k p_k=
\frac{1}{2\pi}\int_{-\pi}^{\pi}dk~p_k\approx
\frac{1}{2\pi\sqrt{2\tau_Q}}.
\label{scaling}
\ee    
The density scales as an inverse of $\hat\xi\propto\tau_Q^{1/2}$, in full consistency with KZM prediction for $\nu=z=1$.
For convenience, we can use the density of kinks to supplement the numerical prefactor
\be 
\hat\xi \equiv \frac{1}{\rho} = 2\pi\sqrt{2\tau_Q},
\label{hatxiprecise}
\ee 
making its inverse equal to the mean density of kinks at the end of the ramp at $g=0$.

%%%%%%%%%%%%%%%%%%%%%%%%%%%%%%%%%%%%%%%%%%%%%%%%%%%%%%%%%%%%%%%%%%%%%%%%%%%%%%%%%%%%%%%%
\subsection{ Ramp to zero transverse field }
\label{app:Weber}
%%%%%%%%%%%%%%%%%%%%%%%%%%%%%%%%%%%%%%%%%%%%%%%%%%%%%%%%%%%%%%%%%%%%%%%%%%%%%%%%%%%%%%%%

To characterize the oscillations, we require more than just the excitation spectrum in Eq.~\eqref{LZpk}. 
A general solution to Eqs.~\eqref{BdGLZ} has the form~\cite{Damski_PRA_2006,QKZteor-e},
\bea
v_k(\tau)&=&
- a D_{-s-1}(-iz) - b D_{-s-1}(iz), \nonumber\\
u_k(\tau)&=&
\left(-\Delta_k\tau+2i\frac{\partial}{\partial\tau}\right)
v_k(\tau),
\label{general} 
\eea
where $D_m(x)$ is a Weber function, 
$s^{-1}=4i\Delta_k$, and $i z=\sqrt{\Delta_k}\tau e^{i\pi/4}$.
Constants $a,b$ are fixed by initial conditions. From the asymptotic behavior of the Weber function when $\tau\to-\infty$, one gets
$a=0$, and 
\be
|b|^2=\frac{e^{-\pi/8\Delta_k}}{4\Delta_k}~.
\ee 
On the other hand, at the end of the ramp when $t=0$ and $\tau=2\tau_Q \sin(2k)$ the argument of the Weber function reads $iz=2\sqrt{\tau_Q}e^{i\pi/4}{\rm sign}(k)\cos(k)$. Its absolute value is large for slow transitions (except near $k=\pm\frac{\pi}{2}$),
and one can again use the asymptotic behavior of the Weber function ~\cite{QKZteor-e}.

When $(t-t_c)\gg \hat t$, the evolution becomes adiabatic, and the time-dependent Bogoliubov modes can be accurately decomposed as
\be 
\left(
\begin{array}{c}
u_k \\
v_k
\end{array}
\right)=
\sqrt{1-p_k}
\left(
\begin{array}{c}
u^+_k \\
v^+_k
\end{array}
\right)+
\sqrt{p_k}
\left(
\begin{array}{c}
v^+_k \\
-u^+_k
\end{array}
\right)
e^{i\varphi_k(t)},
\label{eq:uvafterhatt}
\ee
were $(u^+_k,v^+_k)$ is a positive-frequency stationary Bogoliubov mode at $g(t)$. With the asymptotic behavior of the Weber function, the phase at $t=0$ reads~\cite{QKZteor-e}:
\bea
\varphi_k(0) &=&
\frac{\pi}{4}+
2\tau_Q-
(2-\ln 4)k^2\tau_Q+
k^2 \tau_Q\ln\tau_Q+\nonumber\\
&-&
\arg\left[\Gamma\left(1+ik^2\tau_Q\right)\right]~.
\label{att0smallk}
\eea
Above, $\varphi_k$ is a dynamical phase acquired by a pair of excited quasiparticles $(k, -k)$, with $\Gamma(x)$ being the gamma function. We follow~\cite{RadekNowak}, and approximate ${\rm arg} [\Gamma \left(1+i\tau_Qk^2\right)]\approx -\gamma_E \tau_Qk^2$ to make the phase more tractable. Here $\gamma_E$ is the Euler gamma constant. The approximation is valid for small enough $\tau_Qk^2$, which is consistent with the fact that excited quasiparticles have at most $\tau_Qk^2\approx 1/2\pi$, see Eq.~\eqref{LZpk}. This makes $\varphi_k$ conveniently quadratic in $k$,
\bea
\varphi_k(0) = \frac{\pi}{4} + 2\tau_Q + \left( \ln\tau_Q+\ln 4-2+\gamma_E \right) k^2\tau_Q
\label{eq:quadraticvarphi}
\eea  
This formula completes the full characterization of the Bogoliubov modes when the linear ramp ends at $g=0$.

%%%%%%%%%%%%%%%%%%%%%%%%%%%%%%%%%%%%%%%%%%%%%%%%%%%%%%%%%%%%%%%%%%%%%%%%%%%%%%%%%%%%%%%%
\subsection{ Ramp after $\mathbf{+\hat t}$ }
\label{app:dynamical_phase}
%%%%%%%%%%%%%%%%%%%%%%%%%%%%%%%%%%%%%%%%%%%%%%%%%%%%%%%%%%%%%%%%%%%%%%%%%%%%%%%%%%%%%%%%

In order to obtain phase $\varphi_k(t)$ at earlier time $t<0$, we have to subtract from~\eqref{eq:quadraticvarphi} the dynamical phase acquired between $t$ and the final time when $g=0$. The dynamical phase depends on the spectrum of quasiparticles that, for small $k$, can be approximated as
\be 
\epsilon_k(g) \approx 2(1-g) + \frac{g}{(1-g)} k^2.
\label{epsilonkapprox}
\ee 
The dynamical phase at time $t$ can be obtained as
\bea 
\varphi_k(t) 
&=&
\varphi_k(0) - 2 \int_t^{0} dt' \epsilon_k[g(t')] \\
&=&
\varphi_k(0) - 2 \int_t^{0} dt' \left[ 2[1-g(t')] + \frac{g(t')}{[1-g(t')]} k^2 \right] 
\nonumber\\
&=&
\frac{\pi}{4}+\frac{2t_+^2}{\tau_Q}+
\left(
\gamma_E-\frac{2t_+}{\tau_Q}+\ln\frac{4t_+^2}{\tau_Q}
\right)k^2\tau_Q,
\nonumber
\eea 
where $t_+=t-t_c$. This formula is valid for small $k$ within the support of $p_k$. 

Within the same support we can approximate $u_k^+\approx k/2/(1-g) + {\cal O}(k^3)$ and $v_k^+\approx 1+{\cal O}(k^2)$. With this approximation, relevant products of Bogoliubov modes that follow from~\eqref{eq:uvafterhatt} become
\bea
|v_k|^2 &\approx&
|v_k^+|^2 - p_k - \sqrt{p_k(1-p_k)} \frac{k\cos\varphi_k(t)}{(1-g)} \nonumber\\
u_kv_k^* &\approx&
u_k^+ v_k^+ - p_k \frac{k}{(1-g)} + \sqrt{p_k(1-p_k)} e^{i\varphi_k(t)} . \nonumber
\eea
The last terms with $\varphi_k(t)$ --- that contribute to oscillations --- and the middle terms with $p_k$ are approximated by their leading terms in powers of $k$. They will become the leading terms in $\rho$.

In order to make the following integrals analytically tractable, we approximate~\cite{RadekNowak}:
\bea
\sqrt{p_k(1-p_k)} \approx
e^{-a\pi\tau_Q k^2}
A \sqrt{2 \pi} 
\left(\tau_Qk^2\right)^{1/2}.
\label{A1}
\eea
Here, $A$ and $a$ are variational parameters that can be optimally chosen as $A\approx 19/20$ and $a\approx 4/3$. 
The transverse field is
\be
\langle \sigma^x_n \rangle = 1-2\int_0^\pi \frac{dk}{\pi} \left|v_k\right|^2.
\ee
The integration yields Eq.~(4) in the main text. 
The NN ferromagnetic correlator is
\be 
\langle \sigma^z_n  \sigma^z_{n+1} \rangle = 
2 \int_0^\pi \frac{dk}{\pi} \left|v_k\right|^2 \cos k + 
2 {\rm Re} \int_0^\pi \frac{dk}{\pi} u_k v_k^* \sin k.
\ee 
After approximating $\cos k\approx 1$ and $\sin k\approx k$ to leading order in $k$ and then performing the integral, we obtain Eq.~(7) in the main text that is accurate to leading order in $\rho$.
The other correlator is
\be 
\langle \sigma^y_n  \sigma^y_{n+1} \rangle = 
2 \int_0^\pi \frac{dk}{\pi} \left|v_k\right|^2 \cos k - 
2 {\rm Re} \int_0^\pi \frac{dk}{\pi} u_k v_k^* \sin k.
\ee 
With the same approximations, we obtain 
\be 
\delta^{yy}=
-2\rho-(2-g) \rho^2 d \frac{57\sqrt{6\pi}}{80} \cos \phi.
\label{eq:deltayy}
\ee 
It does not vanish at $g=0$.

%%%%%%%%%%%%%%%%%%%%%%%%%%%%%%%%%%%%%%%%%%%%%%%%%%%%%%%%%%%%%%%%%%%%%%%%%%%%%%%%%%%%%%%%
\section{Oscillations and dephasing after $\mathbf{+\hat t}$}
\label{app:transoscduring}
%%%%%%%%%%%%%%%%%%%%%%%%%%%%%%%%%%%%%%%%%%%%%%%%%%%%%%%%%%%%%%%%%%%%%%%%%%%%%%%%%%%%%%%%

The expansion in Eq.~\eqref{eq:uvafterhatt} becomes accurate at times later than $\hat t$ after the phase transition. 
The phase $\varphi_k$ increases as
\be 
\varphi_k(t)=\int^t dt'~2\epsilon_k[g(t')].
\ee 
After $\hat t$, the quasiparticle spectrum for KZM excitations that are localized near $k=0$ can be considered flat and equal to the gap $\epsilon_0(g)$ that opens with the increasing distance from the critical point. Therefore, the transverse field oscillates with frequency given by twice the instantaneous gap. With~\eqref{epsilonkapprox}
\be 
T=\frac{2\pi}{2\epsilon_0(g)}=\frac{\pi}{2(1-g)}
\label{eq:T}
\ee 
is the period of the oscillations.

Beyond the approximation of flat dispersion, there is some dephasing. The dephasing time can be estimated with the help of the approximate dispersion relation in Eq.~\eqref{epsilonkapprox}. The difference between $\epsilon_k$ for $\hat k\approx1/\sqrt{\tau_Q}$ and $k=0$ is $\delta\epsilon \approx \frac{g}{|1-g|\tau_Q}$. Therefore, the phase gets scrambled on a timescale
\be 
\tau_D = \frac{\pi}{\delta\epsilon} \approx \frac{\pi|1-g|\tau_Q}{g}.
\ee 
When combined with the period~\eqref{eq:T} it yields a quality factor:
\be 
Q = \frac{\tau_D}{T} \approx 2\tau_Q \frac{(1-g)^2}{g}.
\ee 
It diverges at $g=0$ when the dispersion is flat and is the smallest soon after $\hat t$ when $1-g\approx1/\sqrt{\tau_Q}$ and $Q\approx 2$. For a later $g$, it improves with increasing $\tau_Q$, which makes the excitations more narrow in $k$, reducing the dispersion. 

%%%%%%%%%%%%%%%%%%%%%%%%%%%%%%%%%%%%%%%%%%%%%%%%%%%%%%%%%%%%%%%%%%%%%%%%%%%%%%%%%%%%%%%%%%%%%%%%%
\begin{figure*}[t!]
    \includegraphics[width=0.65\columnwidth]{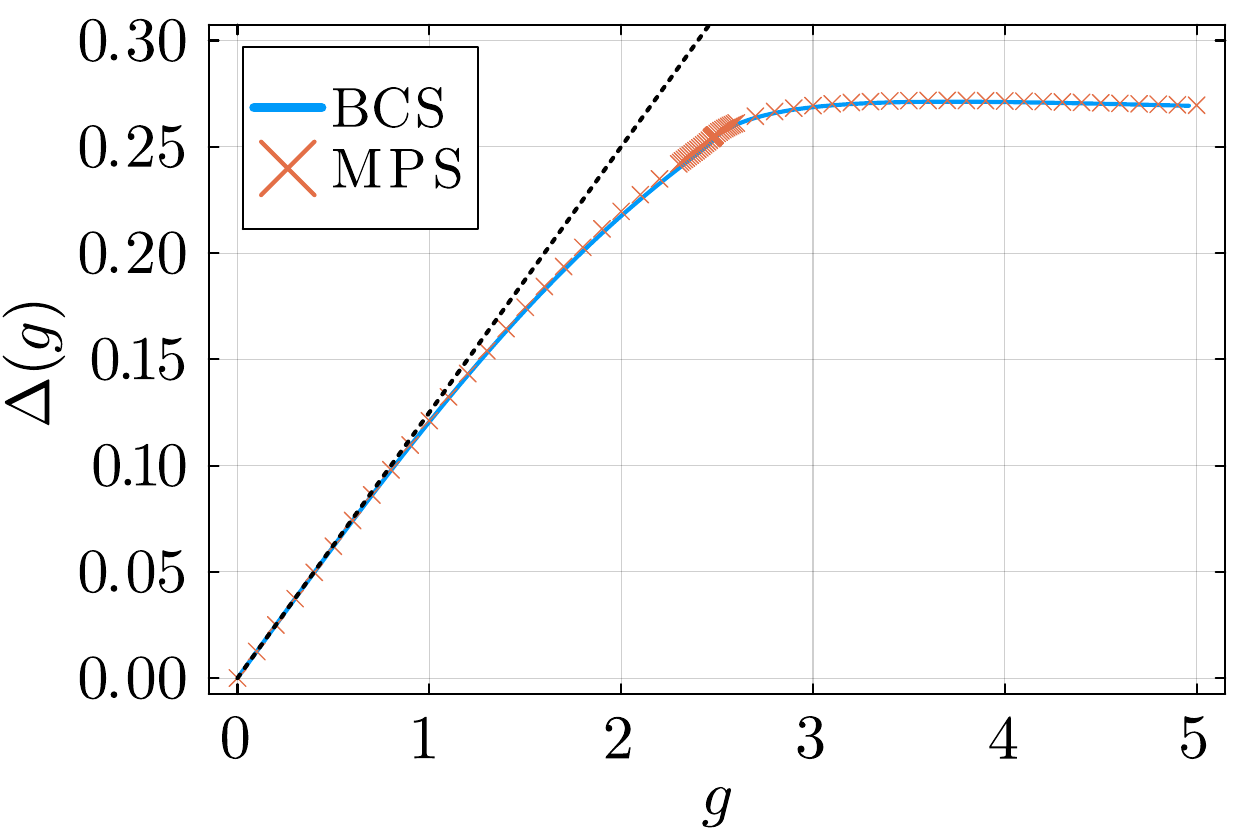}
    \includegraphics[width=0.65\columnwidth]{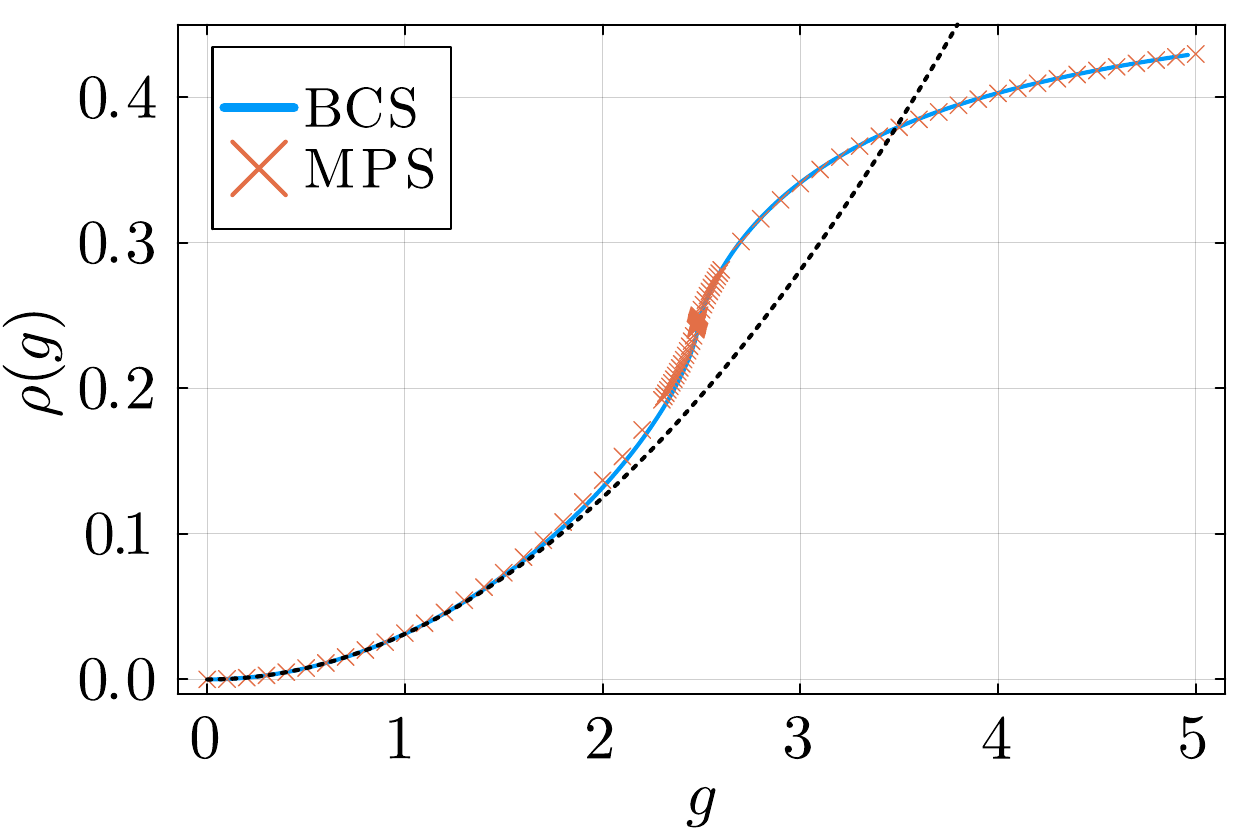}
    \includegraphics[width=0.65\columnwidth]{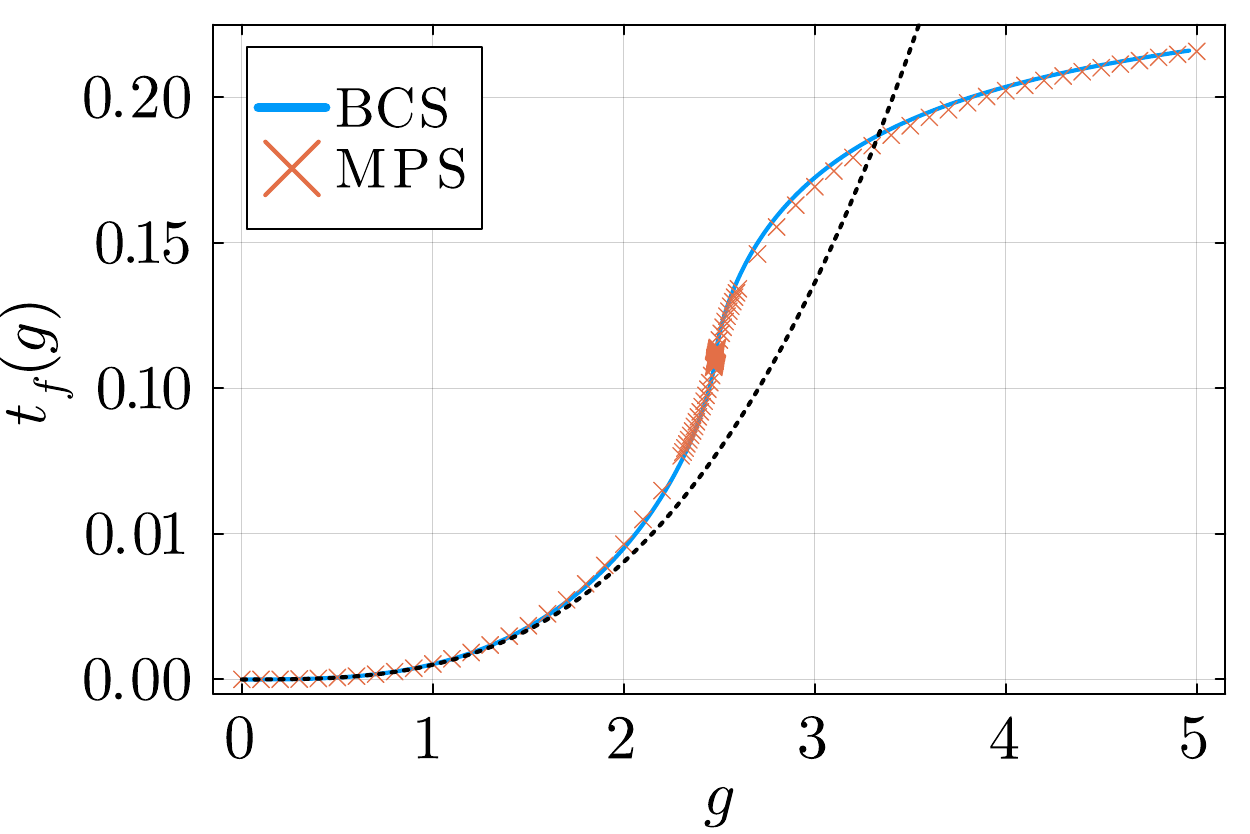}
    \caption{{\bf BCS versus MPS. }
    Comparison between expectation values (from top to bottom) $\Delta$, $\rho$, and $t_f$ obtained within the BCS theory and with infinite matrix product states (MPS). The dotted lines are the perturbative formulas (13) in the main text.% ~\eqref{eq:sss}.
}
    \label{fig:BdG_v_MPS}
\end{figure*}
%%%%%%%%%%%%%%%%%%%%%%%%%%%%%%%%%%%%%%%%%%%%%%%%%%%%%%%%%%%%%%%%%%%%%%%%%%%%%%%%%%%%%%%%%%%%%%%%%%

%%%%%%%%%%%%%%%%%%%%%%%%%%%%%%%%%%%%%%%%%%%%%%%%%%%%%%%%%%%%%%%%%%%%%%%%%%%%%%%%%%%%%%%%
\section{ BCS theory }
\label{app:BCS}
%%%%%%%%%%%%%%%%%%%%%%%%%%%%%%%%%%%%%%%%%%%%%%%%%%%%%%%%%%%%%%%%%%%%%%%%%%%%%%%%%%%%%%%%

\begin{figure}[t!]
    \includegraphics[width=0.49\columnwidth]{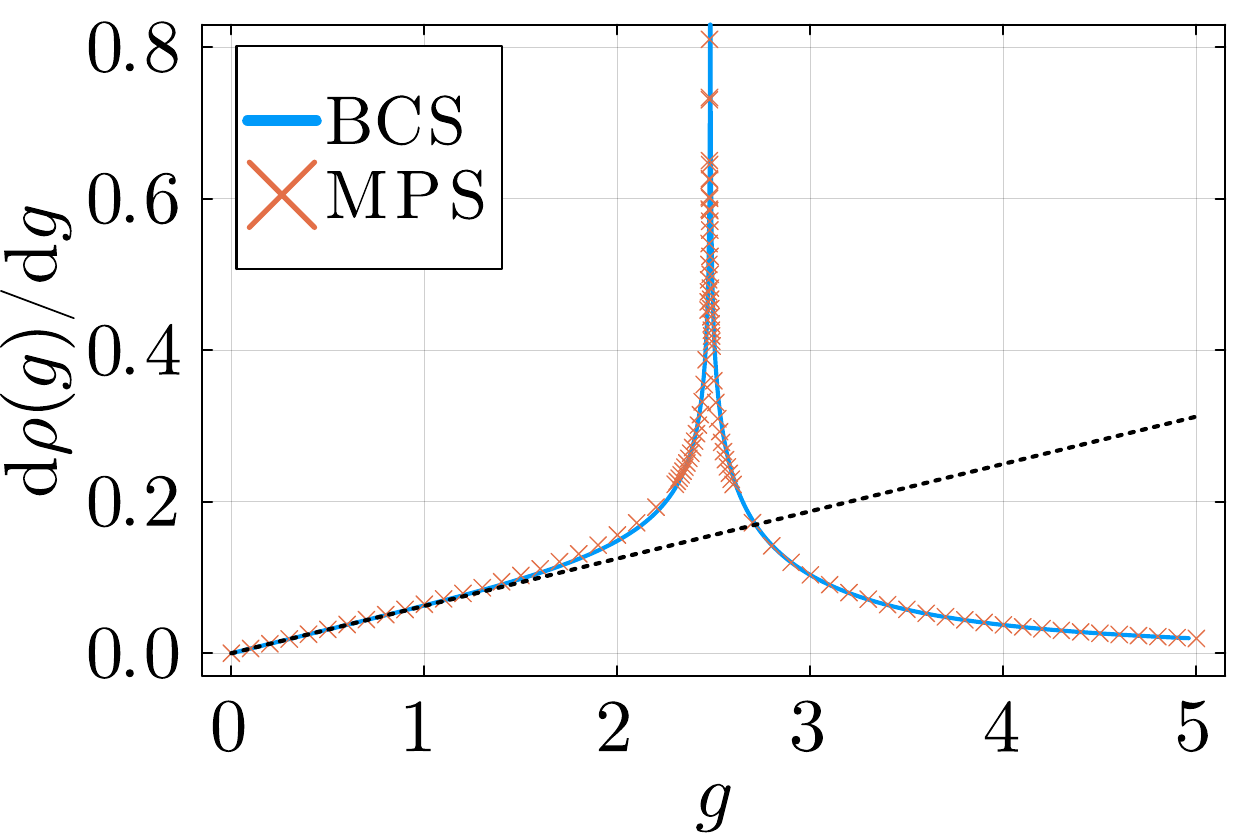}
    \includegraphics[width=0.49\columnwidth]{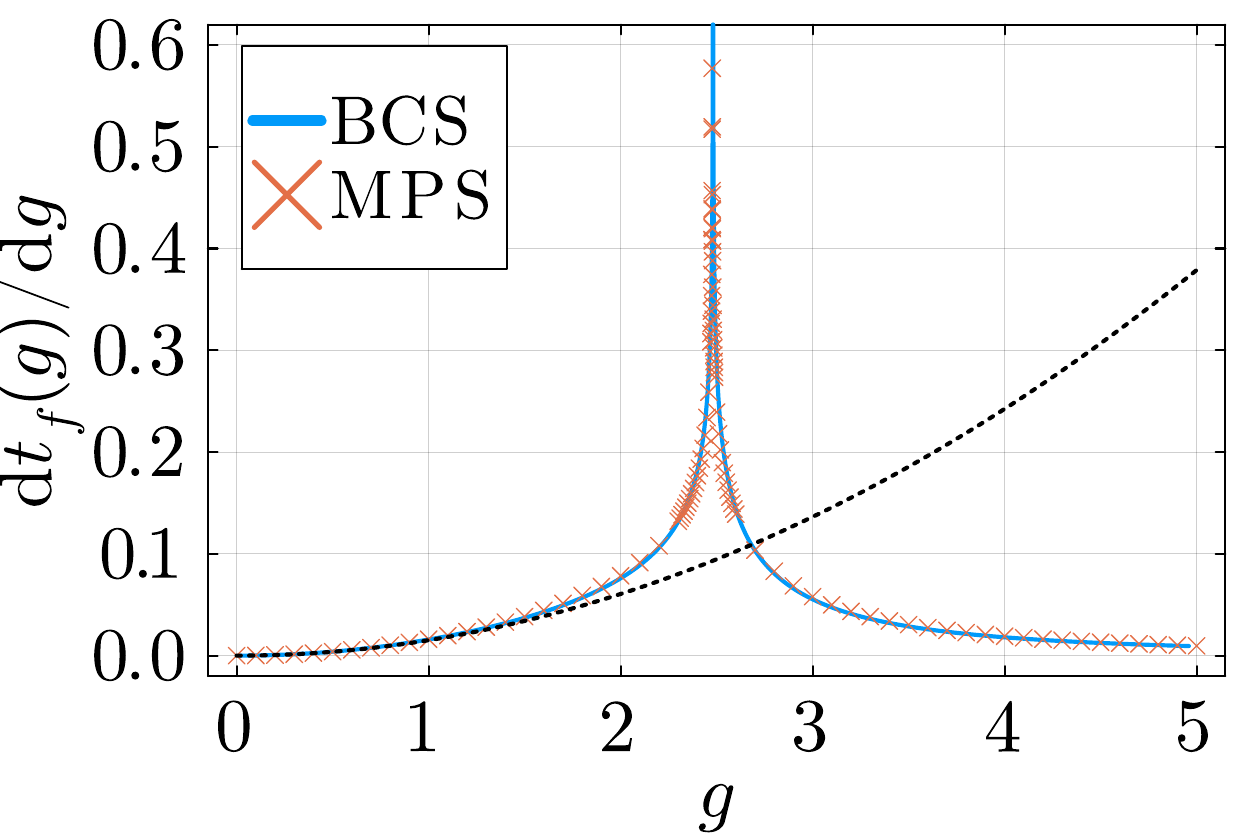}
    \caption{{\bf BCS versus MPS. }
    First derivatives with respect to the field $g$ of $\rho$, and $t_f$ in the ground state from the BCS theory and matrix product states (MPS) in Fig.~\ref{fig:BdG_v_MPS}. A derivative of $\Delta$ is shown in Fig.~2 of the main text. Estimates of $g_{c}$ from BCS and MPS are $g_{c} = 2.48135$ and $g_{c} = 2.47725$, respectively. The dotted lines are the derivatives of the perturbative formulas (13) in the main text.%~\eqref{eq:sss}.
}
    \label{fig:BdG_v_MPS_derivatives_app}
\end{figure}

By Wick's theorem, the energy per site is
\bea 
L^{-1}\bra{0} H \ket{0} 
&=& 
-(J_{1}+J_{2}) + 2\left(J_{1}+2J_{2}\right) \rho \nonumber \\
& &
- g \left(t_f+t_f^*+\Delta+\Delta^*\right) \nonumber \\
& &
- 4 J_2 \left(\rho^2+\Delta^*\Delta-t_f^*t_f\right).
\label{eq:E0}
\eea 
It is energetically favorable we assume $\Delta$ and $t_f$ real. Self-consistency requires the expectation values to satisfy:
\bea 
\rho   &=&   \int_{-\pi}^{\pi} \frac{dk}{2\pi} v_k^2,           \label{eq:rho}\\
t_f    &=&   \int_{-\pi}^{\pi} \frac{dk}{2\pi} v_k^2   \cos k, \label{eq:t}\\
\Delta &=& - \int_{-\pi}^{\pi} \frac{dk}{2\pi} u_k v_k \sin k. \label{eq:Delta}
\eea
Here we used that $u_k,v_k$ are real and have opposite parities with respect to $k$. A self-consistent solution of the Bogoliubov-de Gennes equations is shown in Fig.~\ref{fig:BdG_v_MPS}, where it is compared with results obtained with the matrix product states (MPS). Derivatives of the averages with respect to $g$ are shown in Fig.~2 of the main text and here in Fig.~\ref{fig:BdG_v_MPS_derivatives_app}. The derivatives allow to locate the critical point at $g_{c} = 2.48135$ (BCS) and $g_{c} = 2.47725$ (MPS). These estimates agree within $0.2\%$.

%%%%%%%%%%%%%%%%%%%%%%%%%%%%%%%%%%%%%%%%%%%%%%%%%%%%%%%%%%%%%%%%%%%%%%%%%%%%%%%%%%%%%%%%%%%%%%%%%
%%%%%%%%%%%%%%%%%%%%%%%%%%%%%%%%%%%%%%%%%%%%%%%%%%%%%%%%%%%%%%%%%%%%%%%%%%%%%%%%%%%%%%%%%%%%%%%%%%

%%%%%%%%%%%%%%%%%%%%%%%%%%%%%%%%%%%%%%%%%%%%%%%%%%%%%%%%%%%%%%%%%%%%%%%%%%%%%%%%%%%%%%%%%%%
\section{ Dressed kink quasiparticle operators }
\label{sec:NN_pair}
%%%%%%%%%%%%%%%%%%%%%%%%%%%%%%%%%%%%%%%%%%%%%%%%%%%%%%%%%%%%%%%%%%%%%%%%%%%%%%%%%%%%%%%%%%%

The accuracy of the BCS theory makes it a reliable approximation for the ground state. For $g\ll g_c$, it yields: 
\be 
\rho=\frac{g^2}{32}+{\cal O}(g^4), ~~\Delta=\frac{g}{8}+{\cal O}(g^3), ~~t_f={\cal O}(g^3),
\label{eq:sss}
\ee 
and the Bogoliubov coefficients become
\bea 
u_k &\approx & 
\left(1-\frac{g^2}{64}\right) + \frac{g^2}{64}\cos 2k + {\cal O}(g^3), ~~ \nonumber \\
v_k &\approx & -\frac{g}{4}\sin k -\frac{g^2}{24}\sin 2k + {\cal O}(g^3).
\label{eq:uv_varepsilon}
\eea
With~(11) in the main text, we obtain the kink operator
\bea 
f_n &\approx & 
\left(1-\frac{g^2}{64}\right) \gamma_n 
+ \frac{g^2}{128}\left( \gamma_{n+2} + \gamma_{n-2} \right) \nonumber\\
&&
+ \frac{g}{8} \left(\gamma_{n+1}^\dag-\gamma_{n-1}^\dag\right) +
\frac{g^2}{48} \left(\gamma_{n+2}^\dag-\gamma_{n-2}^\dag\right).
\label{eq:fn}
\eea
Here, the position representation $\gamma_n$ is the Fourier transform~(10) of $\gamma_k$. The $\gamma_n$ is the kink $f_n$ dressed with quantum fluctuations. 
Beyond BCS, the exact Hamiltonian~(9) becomes
\be 
H=E_0+\sum_{k} \omega_k \gamma_k^\dag \gamma_k -4 \sum_n : f_n^\dag f_n f_{n+1}^\dag f_{n+1} :.
\label{eq:Hgamma}
\ee
Here, $E_0=\bra{0}H\ket{0}$, the normal ordering is with respect to $\gamma$, and $\omega_k$ is the quasiparticle dispersion (13) in the main text. With~\eqref{eq:fn} the Hamiltonian~\eqref{eq:Hgamma} becomes 
\bea 
H
& = & 
E_0 +
\omega_\gamma \sum_n \gamma_n^\dag \gamma_n - 
\left(4-\frac{g^2}{8}\right) 
\sum_n \gamma_n^\dag \gamma_{n+1}^\dag \gamma_{n+1} \gamma_n
\nonumber \\
& &
-t_\gamma \sum_n \left( \gamma_n^\dag \gamma_{n+1} + {\rm h.c.} \right)
+\frac{g}{2} \sum_n V_n^{(1)}
\nonumber \\
& &
-t'_\gamma \sum_n \left( \gamma_n^\dag \gamma_{n+2} + {\rm h.c.} \right) + 
\frac{g^2}{16} \sum_n V^{(2)}_n \nonumber\\
& &
+ g^2 \tilde{V}^{(2)} +
{\cal O}(g^3),
\label{eq:Hgamma_approx}
\eea
where
\bea
V_n^{(1)} &=& 
\left( \gamma_{n+2}^\dag \gamma_{n+1}^\dag  + \gamma_{n-1}^\dag \gamma_{n-2}^\dag \right)
\gamma_{n  }^\dag \gamma_{n  }
+{\rm h.c.}, \label{eq:V1} \\
V_n^{(2)} &=&
\left(
\gamma^\dag_{n+1}\gamma^\dag_n+
\gamma^\dag_{n+2}\gamma^\dag_{n+1}
\right)
\gamma_n\gamma_{n-1}+
{\rm h.c.}, \label{eq:V2}
\eea
and 
\bea 
g^2 \tilde{V}^{(2)} &=&
\frac{g^2}{8} \sum_n \gamma_{n+1}^\dag \gamma_{n-1}^\dag \gamma_{n-1} \gamma_{n+1} + \nonumber\\
&&
-\frac{g^2}{16} \sum_n \left( \gamma_{n}\gamma_{n+1}\gamma_{n+2}\gamma_{n+3} + {\rm h.c.} \right) + \nonumber\\
&&
-\frac{g^2}{16} \sum_n \left( \gamma_{n-3}^\dag\gamma_{n}^\dag\gamma_{n}\gamma_{n-1} + {\rm h.c.} \right) + \nonumber\\
&&
-\frac{g^2}{16} \sum_n \left( \gamma_{n+3}^\dag\gamma_{n}^\dag\gamma_{n}\gamma_{n+1} + {\rm h.c.} \right) + \nonumber\\
&&
-\frac{g^2}{6} \sum_n \left( \gamma_{n}^\dag\gamma_{n-1}\gamma_{n}\gamma_{n+1} + {\rm h.c.} \right) + \nonumber\\
&&
-\frac{g^2}{12} \sum_n \left( \gamma_{n}^\dag\gamma_{n}\gamma_{n-1}\gamma_{n-3} + {\rm h.c.} \right) + \nonumber\\
&&
-\frac{g^2}{12} \sum_n \left( \gamma_{n}^\dag\gamma_{n}\gamma_{n+3}\gamma_{n+1} + {\rm h.c.} \right).
\label{eq:tildeV2}
\eea 
collects all remaining second-order terms. For $g\ll g_c$, the first line of~\eqref{eq:Hgamma_approx} dominates, and pairs of NN quasiparticles form tight bound states. These are single reversed spins dressed in quantum fluctuations.

%%%%%%%%%%%%%%%%%%%%%%%%%%%%%%%%%%%%%%%%%%%%%%%%%%%%%%%%%%%%%%%%%%%%%%%%%%%%%%%%%%%%%%%%
\section{ Effective pair Hamiltonian }
\label{app:Heff}
%%%%%%%%%%%%%%%%%%%%%%%%%%%%%%%%%%%%%%%%%%%%%%%%%%%%%%%%%%%%%%%%%%%%%%%%%%%%%%%%%%%%%%%%

The effective pair Hamiltonian, Eq.~(15) in the main text, follows from the kink Hamiltonian \eqref{eq:Hgamma_approx}. It operates in a Hilbert space spanned by Fock states
\be 
b_{n_1}^\dag \dots b_{n_M}^\dag |0\rangle,
\ee
where $n_i+1<n_{i+1}$.
From the second and third term in the first line of Eq.~\eqref{eq:Hgamma_approx}, we obtain a contribution to the pair energy $\omega_b$ in Eq.~(16) of the main text
\be 
\omega_b^{(1)}=2\left(6+\frac{g^2}{8}\right)-\left(4-\frac{g^2}{8}\right)=8+\frac{3g^2}{8}.
\ee 
In terms of pair operators, the third line in Eq.~\eqref{eq:Hgamma_approx} reads
\bea 
&& t'_\gamma \sum_n \left( b_n^\dag b_{n+1} + {\rm h.c.} \right) +\nonumber\\
&& -\frac{g^2}{16} \sum_n \left( b_n^\dag b_{n-1} + b^\dag_{n+1}b_{n-1} + {\rm h.c.} \right).
\eea 
Its contribution to the NN and NNN hopping terms in Eqs.~(15,16) of the main text are
\bea 
t_b^{(1)} &=& -t'_\gamma+\frac{g^2}{16}=-\frac{g^2}{8},\\
t'^{(1)}_b &=& \frac{g^2}{16}.
\eea 
The second line of Eq. \eqref{eq:Hgamma_approx}, which is linear in $g$, breaks/creates/annihilates pairs. Its contributions are second-order perturbative corrections to the pair energy and the NNN hopping. The $t_\gamma$-term contributes:
\bea 
\omega^{(2)}_b &=& -2 \frac{t_\gamma^2}{4} = -\frac{g^2}{2}, \\
t^{(2)}_b      &=&    \frac{t_\gamma^2}{4} =  \frac{g^2}{4},
\eea 
and the $V^{(1)}$-terms contribute:
\bea 
\omega^{(3)}_b &=& -2\frac{(g/2)^2}{4} = -\frac{g^2}{8}, \\
t'^{(3)}_b     &=&   \frac{(g/2)^2}{4} =  \frac{g^2}{16}.
\eea 
Adding up all contributions, we obtain $\omega_b=\omega^{(1)}_b+\omega^{(2)}_b+\omega^{(3)}_b=8-g^2/4$, $t_b=t_b^{(1)}+t_b^{(2)}=g^2/8$, and $t'_b=t'^{(1)}_b+t'^{(3)}_b=g^2/8$.

The second order term \eqref{eq:tildeV2} does not contribute here. Its leading contribution ${\cal O}(g^4)$ could be obtained as a second-order perturbation against quasiparticle energy ($6$ - for terms that change the quasiparticle number) or the pair binding energy ($4$ - for terms that break pairs).

The same $g^2 \tilde{V}^{(2)}$ is not negligible near $g_0$ where the pairs become the lowest branch of excitations and they are not separated by a gap from fermionic quasiparticles. Its third and fourth line drives two fermionic quasiparticles into a pair and back. It is responsible for the population of the pair branch, and the appearance of the oscillations, during the KZ ramp.  

%%%%%%%%%%%%%%%%%%%%%%%%%%%%%%%%%%%%%%%%%%%%%%%%%%%%%%%%%%%%%%%%%%%%%%%%%%%%%%%%%%%%%%%%%%%%%%%%%
\begin{figure}[t!]
    \includegraphics[width=0.49\columnwidth]{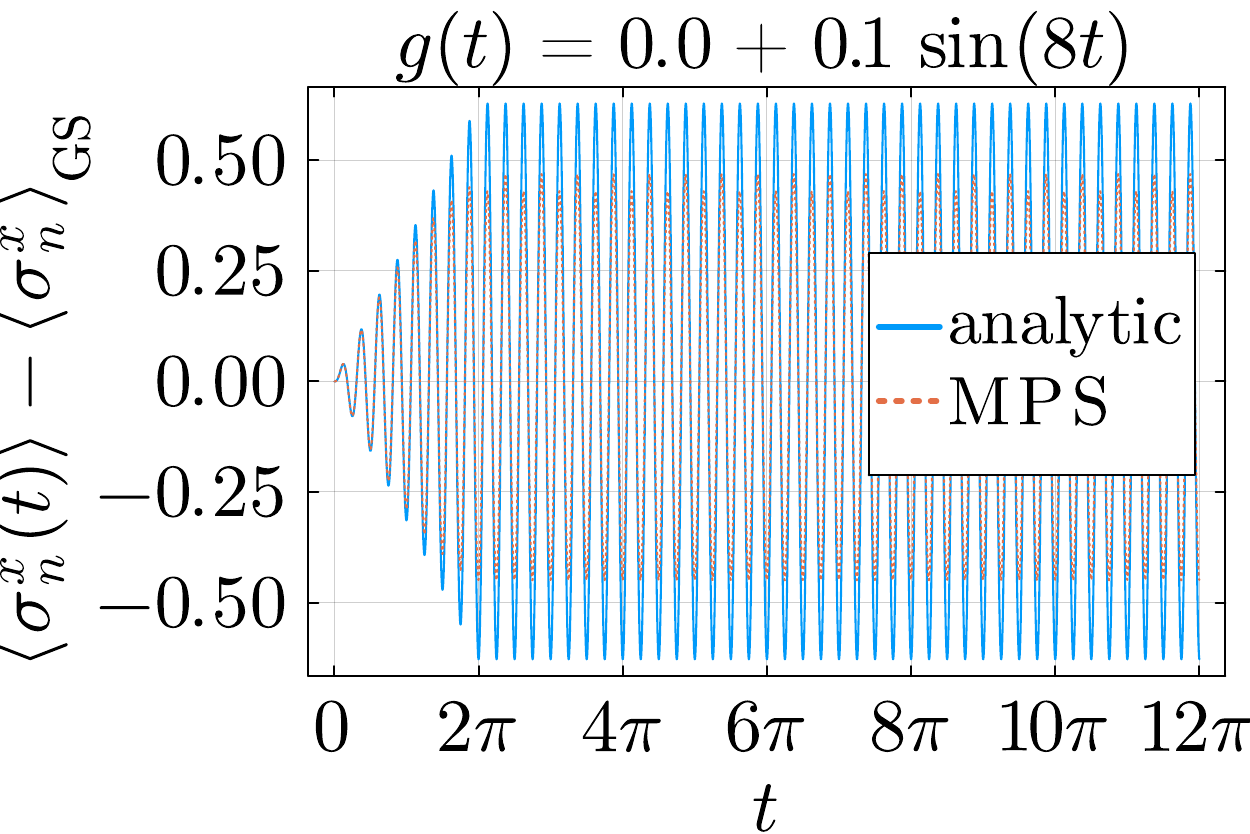}
    \includegraphics[width=0.49\columnwidth]{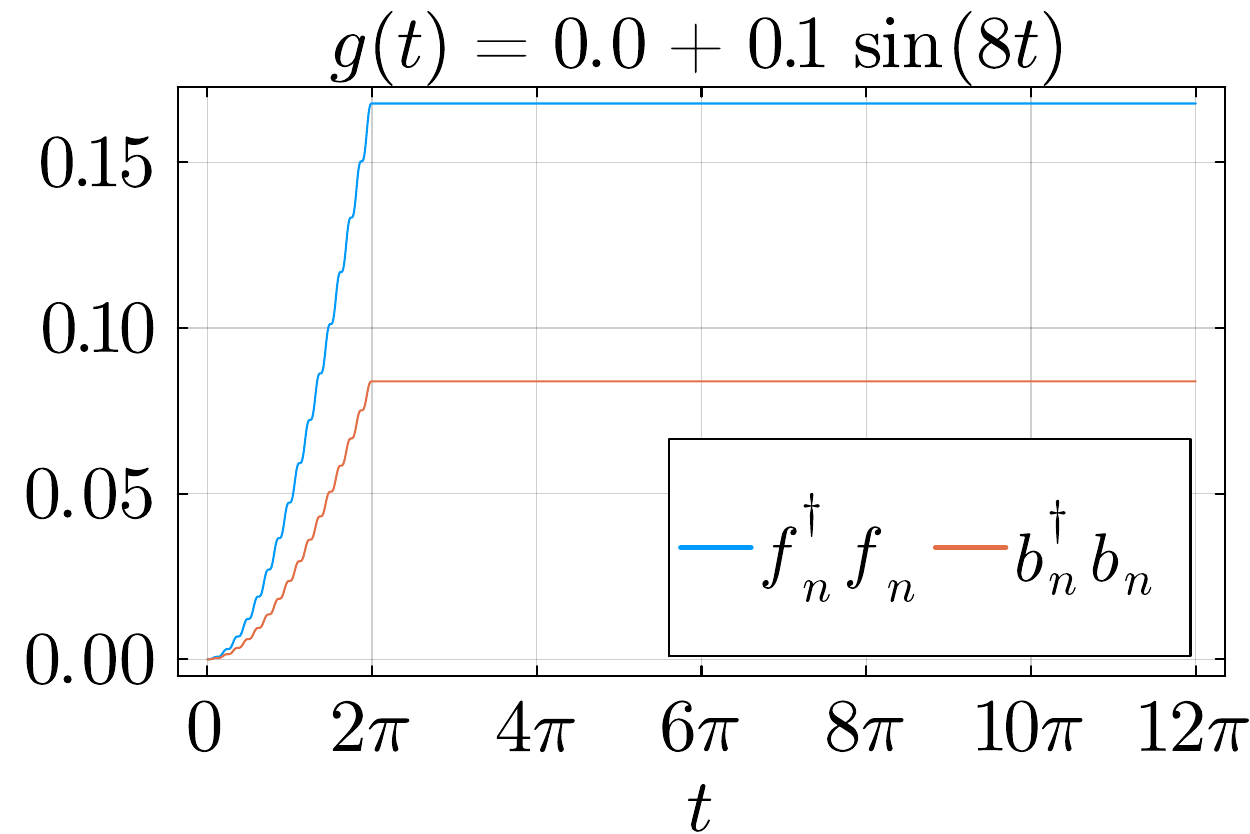}
    \includegraphics[width=0.49\columnwidth]{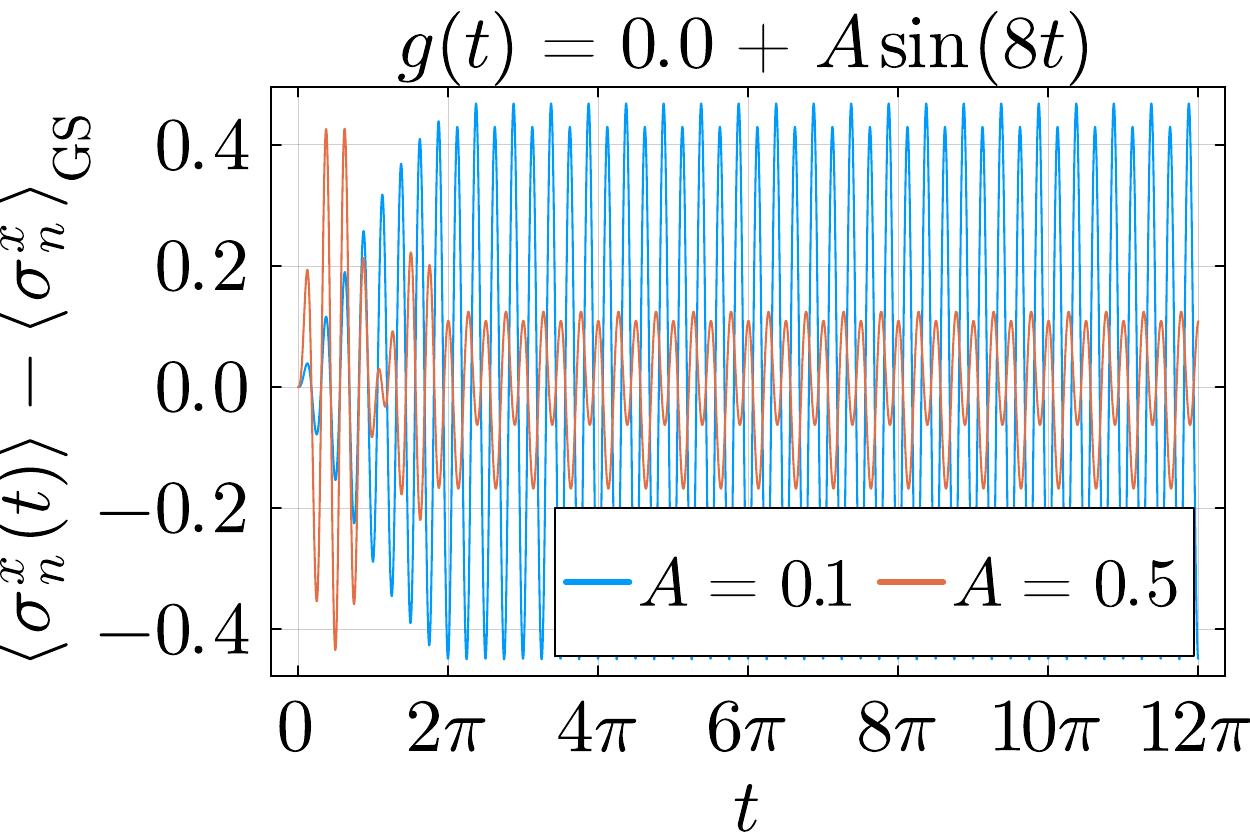}
    \includegraphics[width=0.49\columnwidth]{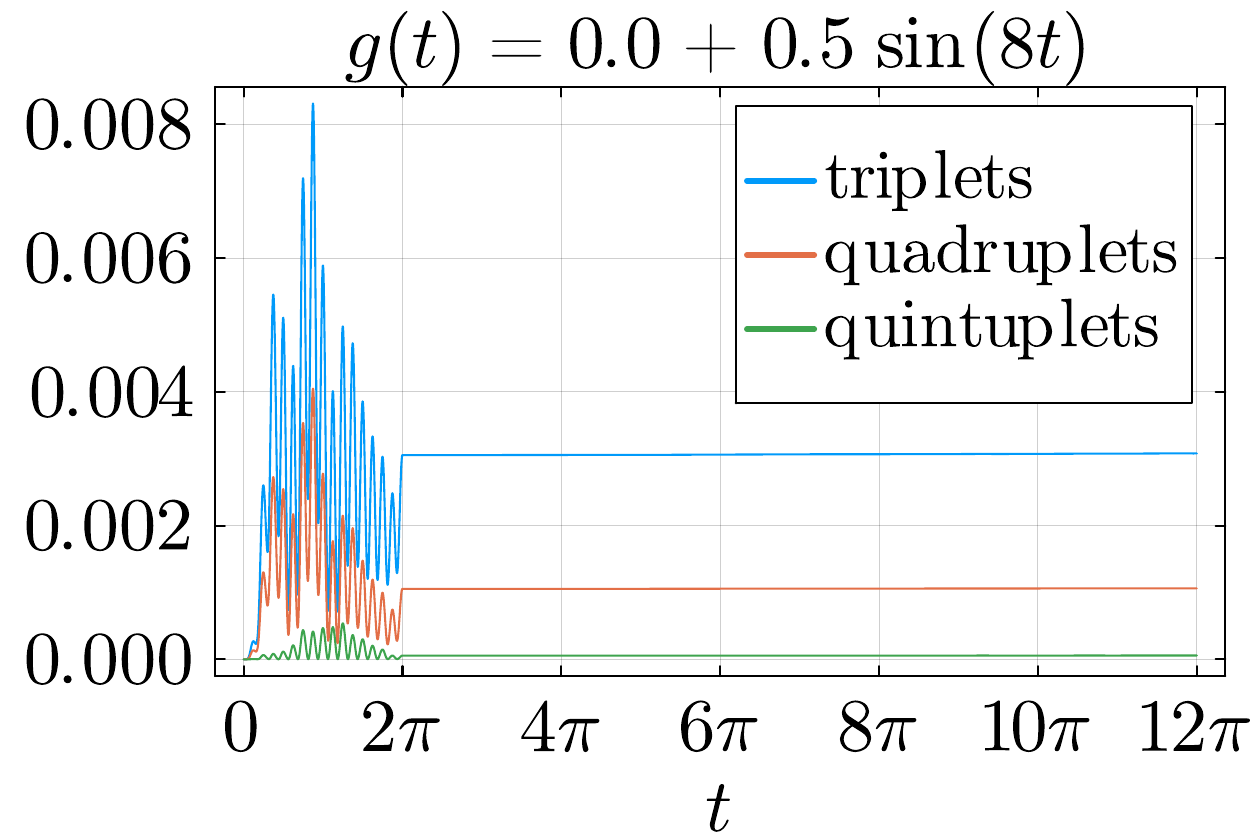}
    \caption{{\bf Crash test at $\mathbf{g=0}$. }
    Top row:
    The periodic driving $\delta g(t)=A\sin(8t)$ at $g=0$ with amplitude $A=0.1$ for time $2\pi$ results in a large amplitude of transverse oscillations. 
    The amplitude is lower than predicted by Eq.~(28) in the main text due to the hard-core nature of bosons $b_n$ that is not quite negligible when their density $0.1$.
    Almost all quasiparticles/kinks are bound into pairs; densities of longer trains are at $10^{-4}$ or lower.
    Bottom row:
    Upon closer inspection, the oscillations for $A=0.1$, and even more $0.5$, have an admixture of frequency $4$ in addition to the main $\omega=8$.
    It is a manifestation of $4$-kink trains with density $0.00008$ and $0.001$, respectively.
}
    \label{fig:test_g0}
\end{figure}
%%%%%%%%%%%%%%%%%%%%%%%%%%%%%%%%%%%%%%%%%%%%%%%%%%%%%%%%%%%%%%%%%%%%%%%%%%%%%%%%%%%%%%%%%%%%%%%%%%

%%%%%%%%%%%%%%%%%%%%%%%%%%%%%%%%%%%%%%%%%%%%%%%%%%%%%%%%%%%%%%%%%%%%%%%%%%%%%%%%%%%%%%%%
\section{Crash test at $\mathbf{g=0}$}
\label{app:crash}
%%%%%%%%%%%%%%%%%%%%%%%%%%%%%%%%%%%%%%%%%%%%%%%%%%%%%%%%%%%%%%%%%%%%%%%%%%%%%%%%%%%%%%%%

Here, we test the theory for strong periodic driving. It is the simplest at $g=0$ when a Bogoliubov quasiparticle is just a kink, 
\be 
\gamma_n=f_n,
\ee 
and the Hamiltonian simplifies to
\bea 
H
& = & 
-L +
6 \sum_n \gamma_n^\dag \gamma_n - 
4 \sum_n \gamma_n^\dag \gamma_{n+1}^\dag \gamma_{n+1} \gamma_n.
\label{eq:Hgamma_approx_g=0}
\eea
The basic theory ignores that the pairs created by $b_n^\dag=\gamma^\dag_{n}\gamma^\dag_{n+1}$ are not exactly bosons. It also ignores trains of kinks/quasiparticles longer than the pair. The trains also make bound states. Their energy at $g=0$ is $6n-4(n-1)=2n+4=6,8,10,12,...$ for $n=1,2,3,4,...$. At nonzero $g$, the terms (19) in the main text can transform a train of $n>2$ quasiparticles into one with $n-2$ --- that has lower energy --- making the longer trains unstable. This mechanism does not exist at $g=0$.

In Fig.~\ref{fig:test_g0}, we drive $\delta g(t)=A\sin(8t)$ with $A=0.1,0.5$ for time $2\pi$. $A=0.1$ is enough to excite large transverse field oscillations that do not quite agree with (28) in the main text. In addition to the lower amplitude of the main frequency $\omega=8$, there is an admixture of frequency $4$. Both effects become stronger for the stronger driving with $A=0.5$, which increases the density of excited kinks. The frequency $4$ originates from a superposition between a train of 4 kinks (energy $12$) and a pair of kinks (energy $8$), e.g.
\be 
\alpha 
\ket{...\uparrow\uparrow\downarrow\uparrow\uparrow\uparrow\uparrow...} e^{-8it} + 
\beta 
\ket{...\uparrow\uparrow\downarrow\uparrow\downarrow\uparrow\uparrow...} e^{-12it},
\ee 
where the corresponding $...$ are the same in both states. The states differ by one reversed spin; hence, $\sigma^x$ has a non-zero matrix element between them, and its expectation value oscillates with frequency $12-8=4$.

%%%%%%%%%%%%%%%%%%%%%%%%%%%%%%%%%%%%%%%%%%%%%%%%%%%%%%%%%%%%%%%%%%%%%%%%%%%%%%%%%%%%%%%%%%%%%%%%%
\begin{figure}[t!]
    \includegraphics[width=\columnwidth]{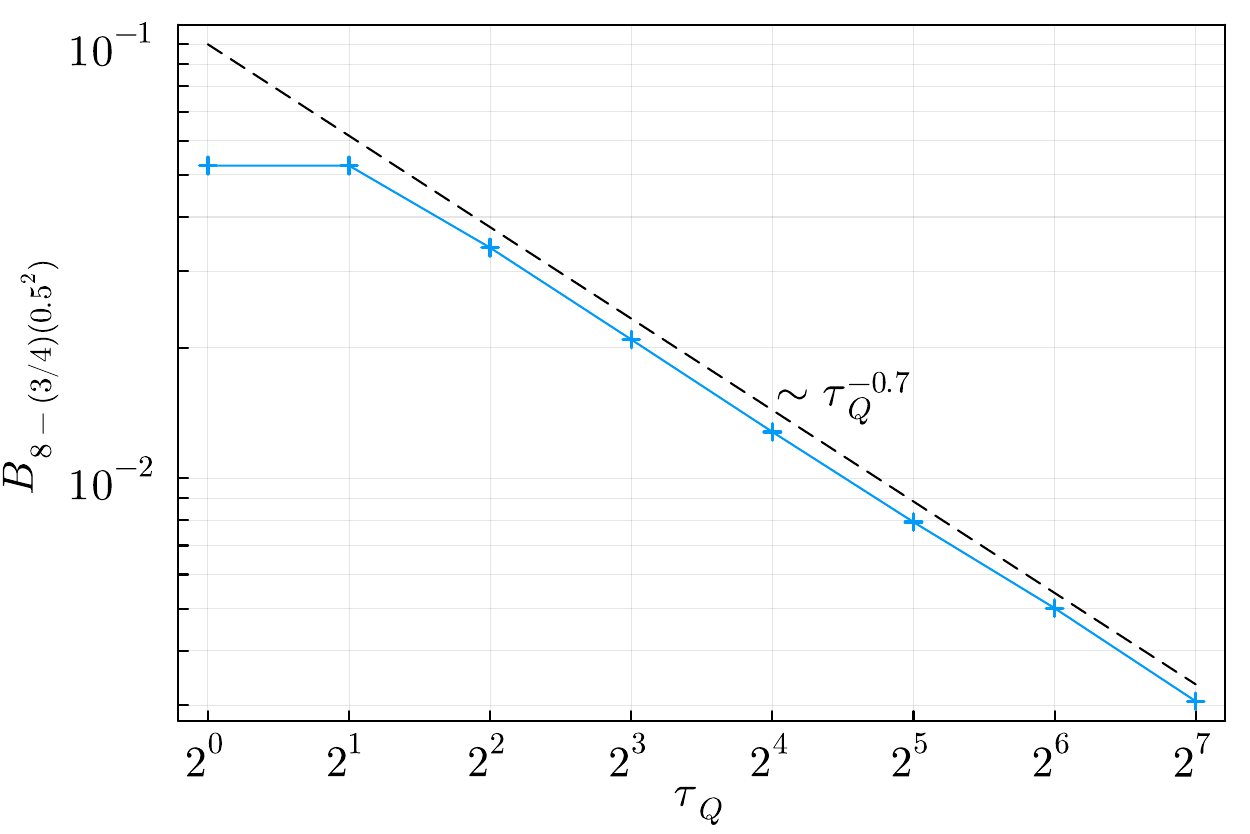}
    \caption{{\bf Amplitude of oscillations. }
    The amplitude of oscillations after a smooth KZ ramp ending at $g=0.5$ with a protocol
    $g(t) = g_{c}[2 - (1-0.5/(2g_{c}))(1+\sin(t/\tau_{Q}))]$, 
    $t\in[-\tau_{Q}\times\pi/2,\tau_{Q}\times\pi/2]$). 
}
    \label{fig:ampl}
\end{figure}
%%%%%%%%%%%%%%%%%%%%%%%%%%%%%%%%%%%%%%%%%%%%%%%%%%%%%%%%%%%%%%%%%%%%%%%%%%%%%%%%%%%%%%%%%%%%%%%%%%

%%%%%%%%%%%%%%%%%%%%%%%%%%%%%%%%%%%%%%%%%%%%%%%%%%%%%%%%%%%%%%%%%%%%%%%%%%%%%%%%%%%%%%%%
\section{ Amplitude of oscillations after KZ ramp}
\label{app:amplitude}
%%%%%%%%%%%%%%%%%%%%%%%%%%%%%%%%%%%%%%%%%%%%%%%%%%%%%%%%%%%%%%%%%%%%%%%%%%%%%%%%%%%%%%%%

After a KZ ramp, the number of kinks $\rho\propto\tau^{-1/2}$ as expected for this universality class ~\cite{transverse_oscillations}. The frequency of the persistent oscillations is consistent with Fig.~3 of the main text. Their amplitude depends on the ramp time with a power law $\tau^{-0.7}$, see Fig.~\ref{fig:ampl}. 

A simple estimate for the exponent is as follows. After crossing the phase transition, the density of excited $\gamma$-quasiparticles is $\rho\propto\tau^{-1/2}$. Near the crossover at $g_0$, when the bound pairs begin to have lower energy, the number of pairs can be roughly estimated as $\rho_p\propto \rho^2\propto \tau^{-1}$. The amplitude is proportional to the square root of $\rho_p$, compare the simple example in Eq.~(23) of the main text, and should scale as $\tau^{-0.5}$. The $0.5$ is close to the $0.7$ but appreciably different. Something is missing in the simple argument. 

The estimate $\rho_p\propto \rho^2\propto \tau^{-1}$ assumes the excited $\gamma$-quasiparticles have no correlations. This is not quite true because they are fermions that must avoid each other at a short distance. In our case, the short distance means less than the characteristic KZ length $\hat\xi$. As the transfer between fermions and pairs is driven by the third and fourth terms in \eqref{eq:tildeV2}, where two fermions need to be separated by exactly 3 sites in order to be transferred into a pair, their short-range repulsion is relevant here. Including the repulsion as e.g. $\rho_p\propto \left(3/\hat\xi\right)^2 \rho^2 \propto \tau^{-2}$ results in an amplitude that decays like $\sqrt{\rho_p} \propto \tau^{-1}$. Here, the power 2 in $\left(3/\hat\xi\right)^2$ is motivated by the analytic formula for the kink-kink correlator in \onlinecite{RadekNowak}. 

%%%%%%%%%%%%%%%%%%%%%%%%%%%%%%%%%%%%%%%%%%%%%%%%%%%%%%%%%%%%%%%%
\begin{figure}[b!]
    \includegraphics[width=0.97\columnwidth]{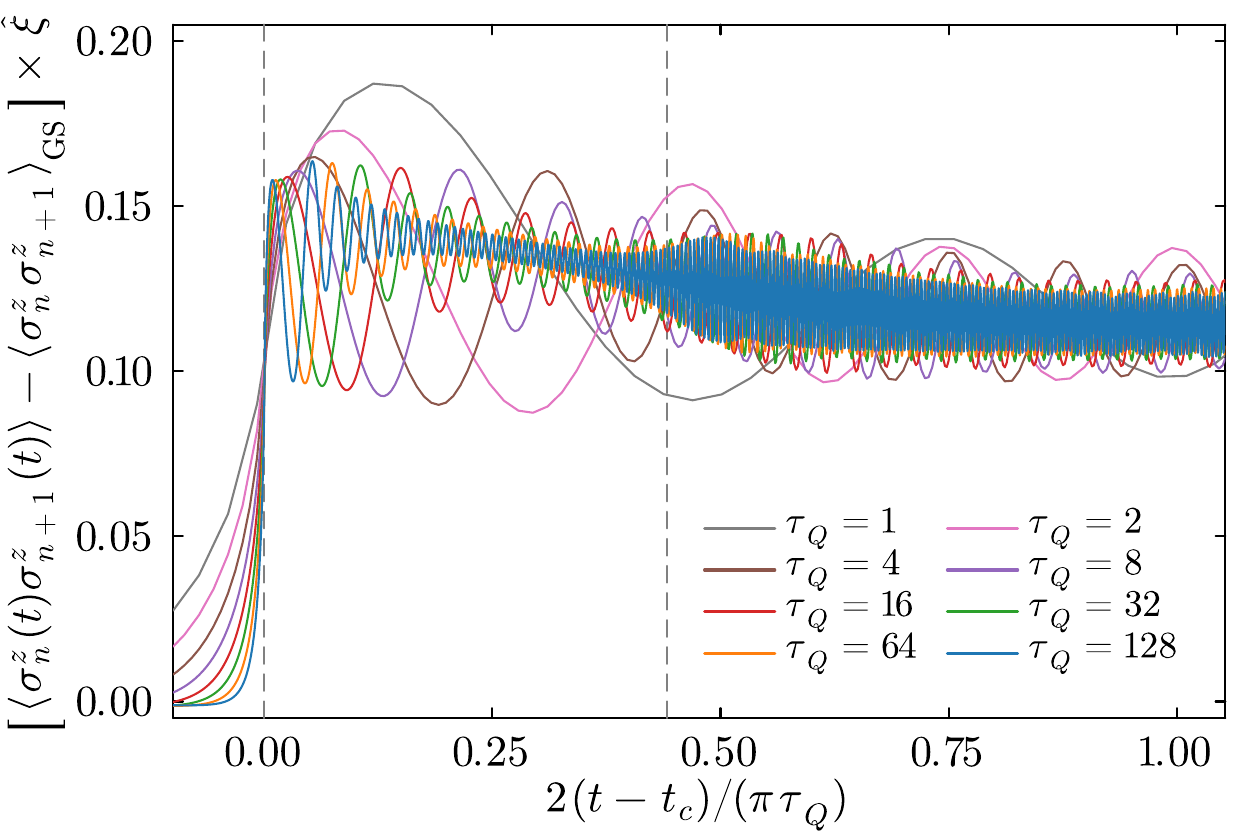}
    \caption{{\bf Persistent coherent oscillation} 
    in the zigzag Ising chain during a quench into the ferromagnetic phase of the rescaled nearest neighbor ferromagnetic correlator:
    $\langle\sigma^z_n\sigma^z_{n+1}\rangle - 
     \langle\sigma^z_n\sigma^z_{n+1}\rangle_{GS}$.
     Similarly to the transverse magnetization in Fig. 1 of the main text, the oscillations become enhanced near $g_0\approx 1.12$ deep in the ferromagnetic phase.
    }
    \label{fig:KZ_ZZ}
\end{figure}
%%%%%%%%%%%%%%%%%%%%%%%%%%%%%%%%%%%%%%%%%%%%%%%%%%%%%%%%%%%%%%%

%%%%%%%%%%%%%%%%%%%%%%%%%%%%%%%%%%%%%%%%%%%%%%%%%%%%%%%%%%%%%%%%%%%%%%%%%%%%%%%%%%%%%%%%
\section{ Oscillations of ferromagnetic energy after KZ ramp }
\label{app:amplitude}
%%%%%%%%%%%%%%%%%%%%%%%%%%%%%%%%%%%%%%%%%%%%%%%%%%%%%%%%%%%%%%%%%%%%%%%%%%%%%%%%%%%%%%%%
The persistent oscillations manifest themselves not only in the transverse magnetization. In some quantum simulation platforms, they may be more accessible via the nearest neighbor ferromagnetic coupling shown in Fig. \ref{fig:KZ_ZZ}.

%%%%%%%%%%%%%%%%%%%%%%%%%%%%%%%%%%%%%%%%%%%%%%%%%%%%%%%%%%%%%%%%%%%%%%%%%%%%

%%%%%%%%%%%%%%%%%%%%%%%%%%%%%%%%%%%%%%%%%%%%%%%%%%%%%%%%%%%%%%%%%%%%%%%%%%%%
\bibliographystyle{apsrev4-2}
\bibliography{KZref.bib}
%%%%%%%%%%%%%%%%%%%%%%%%%%%%%%%%%%%%%%%%%%%%%%%%%%%%%%%%%%%%%%%%%%%%%%%%%%%%

%%%%%%%%%%%%%%%%%%%%%%%%%%%%%%%%%%%%%%%%%%%%%%%%%%%%%%%%%%%%%%%%%%%%%%%%%%%%%%%%%%%%%%%%%%%%%%%%%%%%%%%%%%%%%%